\documentclass[pre,aps,floatfix,showpacs,preprint,superscriptaddress]{revtex4}
\usepackage{graphicx}
\usepackage{verbatim} 
\usepackage{epsfig,graphics,morefloats}
\usepackage{amsmath, amsthm, amssymb}

\newcommand{\be}{\begin{equation}}
\newcommand{\ee}{\end{equation}}
\newcommand{\bea}{\begin{eqnarray}}
\newcommand{\eea}{\end{eqnarray}}

\begin{document}
\title{A Coupled Map Lattice Model for  Rheological Chaos in Sheared Nematic Liquid Crystals}
\author{S. M. Kamil}
\email{kamil@imsc.res.in} 
\affiliation{The Institute of Mathematical Sciences, 
C.I.T. Campus, Taramani, Chennai 600013, India}
\author{Gautam I. Menon}
\email{menon@imsc.res.in} 
\affiliation{The Institute of Mathematical Sciences, 
C.I.T. Campus, Taramani, Chennai 600013, India}
\author{Sudeshna Sinha}
\email{sudeshna@imsc.res.in}
\affiliation{The Institute of Mathematical Sciences, 
C.I.T. Campus, Taramani, Chennai 600013, India}
\affiliation{Indian Institute for Science Research and Education, 
MGSIPAP Complex, Sector 26, Chandigarh 160 019, India}
\date{\today}

\begin{abstract} 
A variety of complex fluids under shear exhibit complex spatio-temporal
behaviour, including what is now termed rheological chaos, at moderate
values of the shear rate. Such chaos associated with rheological response
occurs in regimes where the Reynolds number is very small. It must thus
arise as a consequence of  the coupling of the 
flow to internal structural variables describing the local state of the
fluid. We propose a coupled map lattice (CML) model for such complex
spatio-temporal behaviour in a passively sheared nematic liquid crystal,  
using local maps  constructed so as to accurately describe the spatially 
homogeneous case.  Such local 
maps are coupled diffusively to nearest and next nearest neighbours to mimic
the effects of spatial gradients in the underlying equations of motion.
We investigate  the dynamical steady states obtained as parameters in the map and the strength
of the spatial  coupling are varied, studying  local temporal properties at a single site as well as
spatio-temporal  features of the extended system.  Our methods reproduce
the full range of spatio-temporal behaviour seen in earlier one-dimensional studies
based on partial differential equations. We report
results for both the one and two-dimensional cases, showing that
spatial coupling favours uniform or periodically time-varying states, as intuitively expected. We demonstrate  and
characterize regimes of spatio-temporal intermittency out of which chaos develops.  Our work
suggests that such simplified lattice representations of the spatio-temporal dynamics 
of complex fluids  under shear may provide useful insights as well as  fast and 
numerically tractable  alternatives to continuum  representations.
\end{abstract}
\maketitle
\section{Introduction}
Unusual dynamical steady states are obtained in a large
number of experiments on complex fluids driven out of 
equilibrium\cite{Larson, catesleshouches,safinya,Diat,yanase1991,migler1996}.  When
such fluids are sheared uniformly, the shear stress  $\sigma$ is typically regular
at very small shear rates $\dot{\gamma}$.  However, at larger shear rates the response
is often unsteady, exhibiting oscillations in space and
time as a prelude to intermittency and chaos \cite{sood,soodmore, soodmore1,sood1,Cladis}.
In this non-linear regime, complex fluids under
shear exhibit a variety of instabilities, including instabilities
to  ``shear banded'' states\cite{spenley1993, cates1993,spenley1996,olmsted1997,Olmsted2008, Fielding2007}.  Such banded states  
arise from an underlying multi-valued
constitutive relation connecting the stress and the shear rate, and are
often obtained as a precursor to spatio-temporal intermittency and 
chaotic behaviour in flow response\cite{berret1994, chen1994, berret1995, boltenhagen1997, wheeler 1998, schmidt1999, eiser2000, soubiran2001,salmon}.

Such rheological chaos must be  a consequence of {\em constitutive
non-linearities}, since Reynolds numbers associated with the flow 
are too small for  convective non-linearities to be important\cite{Olmsted2008,Fielding2007}. Such 
constitutive non-linearities originate in the non-trivial internal structure
of the fluid and its coupling to the  flow.  Recent rheological studies of 
``living polymers'' obtain an oscillatory stress response
to steady shear at shear rates above a threshold value\cite{sood,soodmore, soodmore1,sood1}. Such an
oscillatory response turns chaotic at still larger shear
rates\cite{sood,soodmore, soodmore1,sood1}. It has been argued that a hydrodynamic
description of this behaviour requires coupling the internal {\em orientational} state
of such a polymeric fluid to the flow, motivating the study of the 
problem  addressed in this paper\cite{sr,sr1}. This is the  model problem of the
spatio-temporal description of an orientable fluid, such as a nematic 
liquid crystal, placed in a simple steady shear flow\cite{hess30,doi,MD}.

There is a substantial body of previous work on the dynamical states of
complex fluids under shear.   A model due to Fielding and Olmsted 
expresses the stress as a function of a microstructural parameter chosen, for illustrative
purposes, to be the micellar length, which itself evolves in response to the shear rate. The microstructural
parameter yields a viscoelastic contribution to the stress, over and above the
regular fluid contribution\cite{fielding2004}. Fielding and Olmsted show that their model exhibits
spatio-temporal rheochaos. Aradian and Cates have proposed  a one-dimensional model for the instabilities of
a shear-banding fluid system, writing down an  equation
for the time-variation of the shear stress which depends 
both on the instantaneous value of the strain rate  as well as on
the previous history of the stress\cite{aradian2006}. This single non-local equation can be cast as
two coupled local equations, one  for the stress as well as another for a ``memory'' term, 
arising out of the single equation for the stress evolution. This simple
model yields regimes of periodic as well as chaotic behaviour\cite{aradian2006}.

Both these models assume  simplified  scalar descriptions of the internal microstructure. A recent, comprehensive study of a 
shear-banding interface by Fielding and Olmsted, based on the diffusive Johnson-Segalman  (DJS) model, shows that the 
interaction of multiple shear bands can  yield
a time-dependent stress response possessing attributes  of low-dimensional chaos\cite{fielding2006}. However,
such approaches do not examine how such a stress response might arise from an underlying microscopic equation of motion. 
Recent work by Chakraborty, Dasgupta and Sood on a one-dimensional model for nematic rheochaos extends the
model of Refs.~\cite{sr,sr1} by incorporating hydrodynamics, finding stable shear banding as well as 
the coexistence of banded and spatio-temporally
chaotic states\cite{debarshini}. Further, the  DJS model is derivable as a specific limit of their model,
in which the equation for the order-parameter part of the stress is linearized about the isotropic limit. 

In this paper, we present results from a comprehensive study of a simple coupled
map lattice model for rheological chaos, as appropriate to nematic systems under
steady shear. Our local ``microstructural'' variable represents the orientation and degree of 
coarse-grained  order of nematic molecules in the flow, as in the
work of  Refs.~\cite{debarshini,sr,sr1}. We compute the contribution to shear stresses arising
from the evolution of this local variable, showing how uniform, periodic and spatio-temporally
chaotic behaviour in this quantity can be accessed.

The use of coupled map lattices to represent, at a coarse-grained level, behavior
of intrinsically non-linear dynamical systems coupled in space is at least two decades
old\cite{kanekobook}.  Coupled map lattices provide relatively simple models whenever it can be
assumed that the dynamics can be naturally decoupled into a dominant local dynamics representing
behaviour at  a single point in space (or small coarse-grained region) and a spatial coupling term 
which connects this local dynamics weakly across spatial locations. The coupling term idealizes gradient terms in
the underlying continuum equation of motion. Coupled map lattices are well-suited
for computer simulations, since they are naturally discrete in space and time. (Experimental
data are, in fact, close to the CML situation, since any real-life measurement 
requires discrete sampling of the underlying time evolution and
every experiment has some minimum threshold for spatial discrimination, providing a lattice
scale.) Coupled map lattices have been used with success by several authors
in the study of phase-ordering problems as well as in a host of other 
applications\cite{spuri1988, kanekobook}.

We begin by constructing a local map for nematics under 
shear, obtained by discretizing a set of coupled ordinary differential equations (ODE's) describing the continuous time,
spatially local version of this dynamics.  These {\em local} equations have been shown to 
exhibit periodic and regular regimes as well as chaotic regimes. 
We benchmark this map through a detailed comparison to the results from the
study of the ODE system, showing that the qualitative and quantitative aspects of the
phase diagram in this single site limit are rendered accurately. We then 
generalize this to the spatially coupled case by connecting
nearest neighbour maps in a specified manner. The shear enters at the level of the
local map, where it is specified in terms of a single parameter. We take
the point of view that the complexity of the spatio-temporal behavior in the physical
problem can be captured by the most elementary version of spatial coupling,  which, for simplicity 
and following virtually all work on coupled map lattices, we take to be diffusive\cite{kanekobook}.

This {\em local} map  is shown, in agreement with previous work, to exhibit a large number of complex
phases, including uniform (flow aligning in the nematic), tumbling, kayaking and
chaotic phases, in addition to phases which combine one or the other of these
attributes\cite{GR1,GR2}.  While the nematic responds to the fluid through flow alignment as well
as reactive and dissipative terms in the equation of motion, we make the
approximation of ignoring the back-reaction of changes in nematic
order on the fluid. Thus, our approach omits the hydrodynamic interaction, 
since we assume that the flow always remains passive. This is a major assumption. However, it
does have the virtue that a variety of spatio-temporal phenomena with relevance to both the 
experiments as well as to earlier modeling exercises can be demonstrated to
exist in this simple system and are amenable to analysis.

Our second approximation is that we study, for the most part, simple diffusive couplings
between sites, ignoring the advective terms. Consistent with this, we use simple periodic boundary 
conditions on the local field. (We  would otherwise have had to implement a more complex 
Lees-Edwards boundary condition on the fields and ensure an appropriate anchoring
condition at the boundaries\cite{leesedwards}.) Thus, in our model, the shear enters the {\em local}
dynamics but its effects are ignored at larger scales. We pursue this line of investigation
because our interest is specifically in the effects of including spatial couplings into a 
model which provides an accurate description of the temporal behavior of sheared nematics
assuming spatial behavior to be uniform.
We believe, and in some cases have tested this assumption, that incorporating 
the simplest form of spatial coupling should be sufficient for us to be able to explore the full 
spatio-temporal complexity of the sheared nematic problem.

The outline of this paper is the following: Section II outlines our numerical methods
for the construction of the local map. We begin by providing the local equation of motion
for a passively sheared fluid of nematogens, following the work of Refs.~\cite{GR1,GR2}. To 
enforce symmetry and tracelessness, it is customary to project
these (tensor) equations onto a suitable tensor basis. We then construct, through a 
simple Euler discretization, a map within this basis, showing that it can be used to
obtain all the states obtained by ODE-based methods for this problem. The following
section, Section III, describes the construction of the coupled map lattice, illustrating
how the local maps constructed in Section II can be coupled in space, in
both one and two dimensions, along standard lines. Section IV describes our
results in the one-dimensional case, examining the effects of spatial coupling
in both regular and complex regions of the local map. Section V describes
our results for the two-dimensional case, studying, as in the one-dimensional case,
the behaviour in both regular and complex regimes of the local phase diagram.
Section VI contains a discussion of our results as they relate to a quantification of
spatio-temporal complexity in our model, while Section VII contains the
conclusions of this study.

\section{A Local Map for Nematodynamics}

We begin with the continuum equations of motion for a nematic in a specified
flow field. These equations use the tensor representation of the order parameter
in a nematic.  In thermal equilibrium, such order parameter configurations are
weighted by
a Landau-Ginzburg-de Gennes free energy. In a specific Cartesian tensor basis, 
these equations, in the approximation 
that spatial fluctuations in nematic order
are absent, can be cast in terms of equations of motion 
for five expansion coefficients, corresponding to the five independent
parameters characterizing a real symmetric traceless tensor.
These equations of motion, which are ordinary
differential equations (ODE's), are recast as a map, as shown below\cite{footnote}. 

\subsection{Equation of Motion for Nematics}
The derivation of the nonlinear relaxation equations for the symmetric, traceless second rank 
tensor ${\bf Q}$ characterizing local order in a sheared nematic  is available in
earlier work
\cite{hess30,doi,MD,hess36,hess20,MD1,Olmsted,Sheng,Sonnet,
Pleiner,HS}. 
The order parameter is often conveniently expressed as
\be
Q_{\alpha\beta} = \frac{3 s_1}{2} \left ( n_\alpha n_\beta - \frac{1}{3}\delta_{\alpha \beta} \right )
+ \frac{s_2}{2}  \left ( l_\alpha l_\beta -  m_\alpha m_\beta \right ),
\label{definition}
\ee
where the director {\bf n} is defined as the normalized eigenvector corresponding to
the largest eigenvalue of ${\bf Q}$, the subdirector {\bf l} is associated with the sub-leading eigenvalue, and
their mutual  normal {\bf m} is obtained from  {\bf n} $\times$ {\bf l}. The quantities $s_1$ and $s_2$ represent the strength of uniaxial 
 and biaxial ordering: $|s_1| \neq 0$, $s_2=0$ is the uniaxial nematic whereas $s_1,s_2 \neq 0$  with $s_2 < 3s_1$ defines the
biaxial case\cite{degenpro}.

Defining $\widehat{\bf b}: = \frac{1}{2}({\bf b} + {\bf b}^T) - 
\frac{1}{3}(tr{\bf b}) {\bf \delta}$
to be the symmetric-traceless part of the second-rank
tensor {\bf b}, the equation of motion for ${\bf Q}$ in a
velocity field is \cite{hess30,GR2}:
\begin{equation}\label{one}
\frac{d{\bf Q}}{dt} - 2\widehat{{\bf\Omega \cdot Q}} - 
2\sigma' \widehat{{\bf \Gamma \cdot Q}} + 
\tau_Q^{-1}{\bf \Phi} = - \sqrt{2}\frac{\tau_{ap}}{\tau_a}{\bf\Gamma},
\end{equation}
where the tensor 
${\bf \Omega} = \frac{1}{2}((\nabla{\bf v})^T - \nabla{\bf v})$,
${\bf \Gamma} = \frac{1}{2}((\nabla{\bf v})^T + \nabla{\bf v})$
and $\nabla{\bf v}$ is the velocity gradient tensor, 
with ${\bf v} = \dot{\gamma}y{\bf e^x}$, where
${\bf e^x}$ is a unit vector in the $x-$ direction.
The velocity is along the $x$ direction,
the velocity gradient is along the $y$ direction, while
$z$ is the vorticity direction.
The quantities
$\tau_a > 0$ and $\tau_{ap}$ are phenomenological relaxation times,
$\sigma'$ describes the change of alignment 
caused by ${\bf \Gamma}$ and ${\bf \Phi} = \partial \phi/\partial{\bf Q}$,
with the free energy $\phi({\bf Q})$ given by
 \be
\phi({\bf Q}) = \frac{1}{2}A{\bf Q:Q} - 
\frac{1}{3}\sqrt{6}B({\bf Q \cdot Q}):{\bf Q} + \frac{1}{4}C({\bf Q:Q})^2.
\label{free}
\ee
If the spatial variation is also taken into account, $\nabla^2{\bf Q} $ and $\nabla \nabla \cdot{\bf Q}$
should also be included in the above expression . The notation $Q:Q$ represents $Q_{ij}Q_{ji}$, with repeated indices
summed over.
Here $A = A_0(1-T^*/T)$, and $B$ and $C$ are constrained by the 
conditions $A_0 > 0$, $B>0,C>0$ and $B^2 >\frac{9}{2}A_0C$. 

The symmetric traceless alignment tensor {\bf Q} has five independent components.
Assuming spatial uniformity, so that gradients of the  ${\bf Q}$ tensor can be dropped, 
a system of 5 coupled ordinary differential equations (ODEs) for the 5 independent
components of ${\bf Q}$ can be obtained with the choice of a suitable tensor basis.  Choosing
the standard orthonormalized Cartesian tensor basis leads to the expansion
\begin{equation}
{\bf Q} = \sum_{k = 0}^4 a_k {\bf T^k},
\end{equation}
with
\begin{equation}
\quad {\bf T^0}= \sqrt{3/2} \widehat{{\bf e}^z{\bf e}^z},\quad  {\bf T^1} =\sqrt{1/2}({\bf e}^x{\bf e}^x - {\bf e^y}{\bf e^y}),\nonumber \\
\end{equation}
\begin{equation}
 {\bf T^2}= \sqrt{2} \widehat{{\bf e}^x{\bf e}^y},\quad  {\bf T^3}= \sqrt{2} \widehat{{\bf e}^x{\bf e}^z},\quad  {\bf T^4}= \sqrt{2} \widehat{{\bf e}^y{\bf e}^z}.
\end{equation}
\subsection{Dynamics of Sheared Nematics from a Local Map}

We work in the tensor basis described above, representing the equations of motion of
Eq.~\ref{one} in terms of the coupled equations of motion for the five coefficients $a_0 \ldots a_4$.  
The problem of representing 
the time updates in terms of a local map is most easily approached by considering
the lowest order Euler discretization of the underlying differential equations. (There are alternative
methods of constructing maps from local dynamics governed by ODE's, including stroboscopic
methods and methods which use Poincare sections; however, the choice
we have made is the simplest given the variety and complexity of the dynamical states
we would like to describe.)

Scaling parameters as in Ref.~\cite{GR1,GR2}, and making the same choice of
numerical values  as in Ref.~\cite{sr,sr1}, we obtain the following map
\begin{eqnarray}
\label{local}
f_0(a_0^{t}) &= {a_0}^t + &\Delta \left(-(2 a^{2} - 3 a_{0})a_{0} - 3(a_{1}^{2} 
              + a_{2}^{2}) + \frac{3}{2}(a_{3}^{2} +a_{4}^{2}) 
							\right)^t
		\nonumber \\
	f_1({a_1}^{t}) &= {a_1}^t + &\Delta \left(-(2 a^{2} + 6 a_{0})a_{1} 
	                  + \dot{\gamma}a_{2} + \frac{3}{2}\sqrt{3}(a_{3}^{2} 
										- a_{4}^{2})\right)^t
	  \nonumber \\
	f_2({a_2}^{t}) & = {a_2}^t + &\Delta \left(-(2 a^{2}+ 6 a_{0})a_{2} 
	                    - \dot{\gamma}a_{1} + 3\sqrt{3}a_{3}a_{4} 
											+ \frac{\sqrt{3}}{2}\lambda_{k}\dot{\gamma} 
											\right)^t 
											\nonumber\\
f_3({a_3}^{t})& = {a_3}^t + &\Delta \Bigg(-(2 a^{2}- 3 a_{0})a_{3} 
                  + \frac{1}{2}\dot{\gamma}a_{4} + 3\sqrt{3}(a_{1}a_{3} 
									+ a_{2}a_{4})\Bigg)^t
\nonumber \\
f_4({a_4}^{t}) & = a_4^t + &\Delta \Bigg(-(2 a^{2}- 3 a_{0})a_{4} 
                   - \frac{1}{2}\dot{\gamma}a_{3} + 3\sqrt{3}(a_{2}a_{3} 
									 - a_{1}a_{4})\Bigg)^t.
\end{eqnarray}
Here $t$ indicates discrete time steps, and $\{{.} \}^t$ denotes the
value of the quantity $\{{.}\}$ at time step $t$. All the functions
$f$ denote the locally updated value of their argument at a time step
($t+1$). For the purely local map, $f_i(a_i^{t}) \equiv a_i^{t+1}$; however,
for the coupled map, the value of $f_i(a_i^{t})$ is computed as an
intermediate step, prior to the diffusive step which yields
the final quantity $a_i^{t+1}$.

The quantity $a = a_0^2 +a_1^2 +a_2^2 +a_3^2 +a_4^2$.
We choose $\Delta = 0.01$ for all our calculations,
but have checked that changing $\Delta$ by upto an order of magnitude does not
affect our results.
Our choice of parameters implies that the system in the
absence of shear is at the limit of metastability of the isotropic phase. Our
choice of the value for $\Delta$  captures all the features of
 the full local phase diagram obtained in Refs. \cite{GR1,GR2}. 

The order parameter part of the stress is proportional to contributions from the
Landau-de Gennes free energy as well as from the
gradient terms, which we represent through the spatial coupling term in the coupled
map lattice. This is obtained as described in the following sections.

\subsection{Phase Behaviour of the Local Map}

\begin{figure}[!h]
\begin{center}
\includegraphics[width=3.0in]{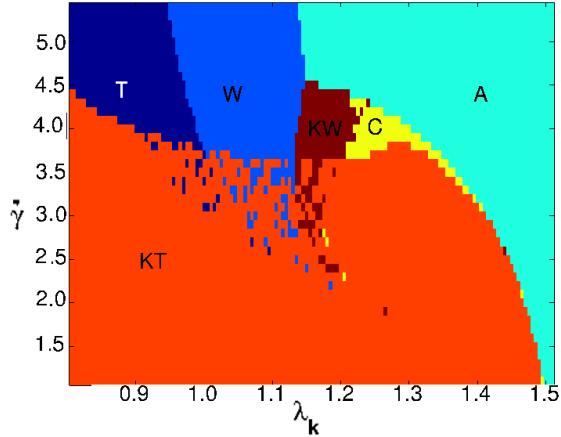}
\caption{[Color Online] Phase diagram for dynamical behaviour in the local map defined 
through Eqns.\ref{local}, with  the parameter $\lambda_k$ plotted
on the $x$ axis and  $\dot{\gamma}$ on the $y$ axis. 
Here T denotes the tumbling state, W  the wagging state, KT  the  kayak-tumbling, 
KW the kayak-wagging state, A the aligned state and C the state in which complex 
dynamics is seen. These states are discussed further in the text. }
\label{local_phase2}
\end{center}
\end{figure}

Examining the dynamical steady states of this map at a large number of points in
the space spanned by $(\dot{\gamma}, \lambda_k)$ 
yields a complex phase diagram admitting many states -- aligned, tumbling, wagging,
kayak-wagging, kayak-tumbling and chaotic -- as functions of the shear
rate $\dot{\gamma}$ and a phenomenological relaxation time $ \lambda_k$ which is a
parameter in the equations of motion\cite{GR1,GR2,Grosso}. Fig.~(\ref{local_phase2}) exhibits the dynamical states found in the
map for the uncoupled case, in terms of a phase diagram in the
quantities $\lambda_k$ and $\dot{\gamma}$. Such a phase diagram bears
considerable similarities to phase diagrams obtained by other authors in the
PDE representation;  see, for example, Fig. ~7 of 
Ref.~\cite{GR2}.

The states in this phase diagram are labeled as  follows: The first is the
state labeled $A$, which is the {\bf Aligned state},
where all  dynamics ceases, and the director is aligned at an angle to the
flow. In the standard Couette geometry, the
velocity field and the velocity gradient form a plane, called the
vorticity plane. In our case, this is the $x-y$ plane. If the director lies in the vorticity plane and
rotates about an axis (the $z-$axis) perpendicular to this plane, the dynamical state
is called a {\bf Tumbling state}. The tumbling state is denoted by
$T$ in the phase diagram of Fig.~(\ref{local_phase2}). If
the director, while lying in the plane, executes oscillations, the
dynamical state is called a {\bf Wagging state}. The wagging state is
represented in the local phase diagram by the symbol $W$. 

If the director rotates and oscillates,  moving out of  the vorticity
plane, the dynamical states are called {\bf Kayak-Tumbling} and
{\bf Kayak-Wagging} respectively. They are represented as KT and KW in
the local phase diagram. If the dynamics is a mixture  of
complex intermittent behaviour and coexisting attractors, the state is called {\bf Complex} and
is represented by $C$ in the phase diagram. Clearly the interesting
region in the phase diagram lies in and near the region labeled C.

Fig.~\ref{local_phase2} is obtained in the following way. The phase-space
of the $\dot{\gamma}$ and $\lambda_k$ variables is gridded and an initial
random initial condition chosen at each point. After the passage of an initial
transient state, the system
goes to dynamical attractors, ranging from simple spatiotemporal fixed points to 
complex intermittent behaviour.
These dynamical attractors are  identified with one of the states described above, i.e.
$A,K,T,KW,KT$ or $C$. In some regimes, one sees a coexistence
of states {\it i.e.} KT and T and KW and W {\it i.e.} different initial conditions
can give rise to different asymptotic behaviour in the long time limit.

\section{Coupled Maps for Nematodynamics}
Our spatially coupled model is built up from the  local maps 
given in  Eq.~\ref{local}. 
These maps are  placed on the sites of a regular lattice in one and two dimensions 
and  can be coupled  via  several   different coupling schemes, as described below. The generalization to
arbitrary dimensions as well as different coupling schemes is a straightforward 
one. 

For a one dimensional lattice,  with  sites  indexed by  the label $i$, the
five variables ($a_0 (i) \dots a_4 (i)$) on each lattice site evolve
in discrete time $t$ as:
\be
\label{one2}
\phi_i =\phi'_i + \frac{1}{3}\epsilon \,\big( \phi^t_{i+1}+\phi^t_{i-1}
                    - 2\phi^t_i \big),
\ee
where $\phi \in (a^{t+1}_0,\, a_1^{t+1},\,a^{t+1}_2,\, a_3^{t+1},\,
a_4^{t+1})$ and $\phi' \in (f_0(a_0^t) ,f_1(a_1^t) ,f_2(a_2^t)
,f_3(a_3^t) ,f_4(a_4^t)) \;. Here \epsilon$ is a coupling constant which is chosen to take
values between $0$ and $3/2$.

For the two dimensional case we consider a square lattice with site
index $(i,j)$ \; and with the set of five variables $(a_0(i,j),\;
a_1(i,j),\; a_2(i,j),\; a_3(i,j),\; a_4(i,j))$ on each lattice point at
time-step $t$ evolving in time as :
\begin{eqnarray}
\label{2}
\phi_{i,j}& =\phi'_{i,j}& + \frac{1}{6}\epsilon \,\big( \phi^t_{i+1,j}+\phi^t_{i-1,j}
                    +\phi^t_{i,j+1}+\phi^t_{i,j-1} \big) \nonumber\\
				      &\qquad &+\frac{1}{12}\epsilon \,\big( \phi^t_{i+1,j+1}+\phi^t_{i-1,j-1}
							 + \phi^t_{i-1,j+1}+\phi^t_{i+1,j-1} \big)- \epsilon \phi^t_{i,j},
\end{eqnarray}
where $\phi \in (a^{t+1}_0,\, a_1^{t+1},\,a^{t+1}_2,\, a_3^{t+1},\,
a_4^{t+1})$ and $\phi' \in (f_0(a_0^t) ,f_1(a_1^t) ,f_2(a_2^t)
,f_3(a_3^t) ,f_4(a_4^t)) \;, $ and $\epsilon$ is a coupling constant having
value between $0$ and $1$.
The choice
of the numerical coefficients $1/6$ and $1/12$ in the coefficients of the
nearest and next-nearest neighbour terms are standard choices
in the CML literature. They represent choices of  lattice  discretization which
are as close as possible to the continuum limit.

The local value of the shear stress $(\sigma_{xy})_{i,j}$  at the (two-dimensional)
site $(i,j)$ is obtained from the following definition\cite{footnote1}:
\begin{eqnarray}
\label{stress}
(\sigma_{xy})_{i,j}& =(\sigma'_{xy})_{i,j}& + \frac{\sqrt{2}}{6}\epsilon \,\big( \phi_{i+1,j}+\phi_{i-1,j}
                    +\phi_{i,j+1}+\phi_{i,j-1} \big) \nonumber\\
				      &\qquad &+\frac{\sqrt{2}}{12}\epsilon \,\big( \phi_{i+1,j+1}+\phi_{i-1,j-1}
							 + \phi_{i-1,j+1}+\phi_{i+1,j-1} \big)-\sqrt{2} \epsilon \phi_{i,j},
\end{eqnarray}
where $(\sigma'_{xy})_{i,j}$ is given by
\be
(\sigma'_{xy})_{i,j} = (2 \sqrt{2}a_2 a^2-6\sqrt{6}(a_3a_4/2-(a_2a_0/\sqrt{3})))_{i,j},
\ee
and $\phi = a_2$.

The definition for the one-dimensional case follows from:
\begin{eqnarray}
\label{stress_oned}
(\sigma_{xy})_i& =(\sigma'_{xy})_i& + \frac{\sqrt{2}}{3}\epsilon \,\big( \phi_{i+1}+\phi_{i-1} - 2\phi_i \big),
\end{eqnarray}
where $(\sigma'_{xy})_i$ is given by
\be
(\sigma'_{xy})_i = (2 \sqrt{2}a_2 a^2-6\sqrt{6}(a_3a_4/2-(a_2a_0/\sqrt{3})))_i,
\ee
and $\phi = a_2$.

We have also experimented with other choices of the update rule. While the
update rule of Eq.~\ref{one2} can be termed as the {\em post-update} rule, in which
the terms on the right hand side are calculated at time $t+1$, one could
alternatively use the {\em pre-update} rule, in which these terms are calculated
at time $t$. We have checked that varying this choice of update rule does not
affect our results. In the equation for the two-dimensional update, (Eq.~\ref{2}), we
have checked that dropping the next-nearest neighbour term also does not
affect our results significantly. Thus, a variety of possible update schemes appear
to yield consistent results for the spatio-temporal behaviour of our coupled map
lattice, underlining the generic nature of our results.

In the sections below we describe the relevant features emerging from
the simulation of the coupled map lattice above for a wide range of
parameters and initial states.

\section{The One-dimensional Coupled Map Lattice}
In this section, we describe our results for the one-dimensional case, concentrating on
the effects of the inter-site coupling, both within and outside the regime labelled C
in the phase diagram.
\subsection{Local dynamics}
 \begin{figure}
\begin{center}
\includegraphics[width=3.0in]{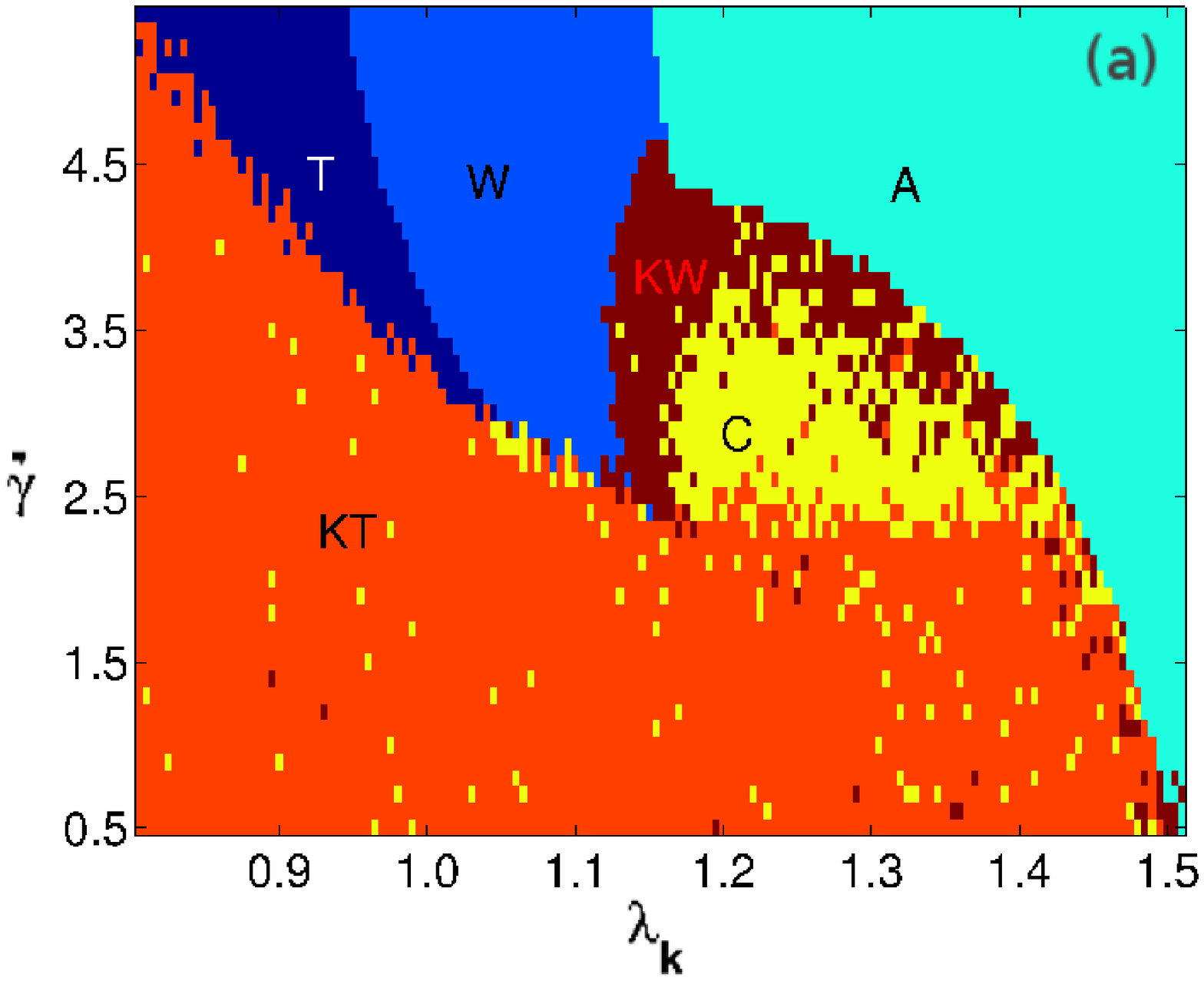}
\includegraphics[width=3.0in]{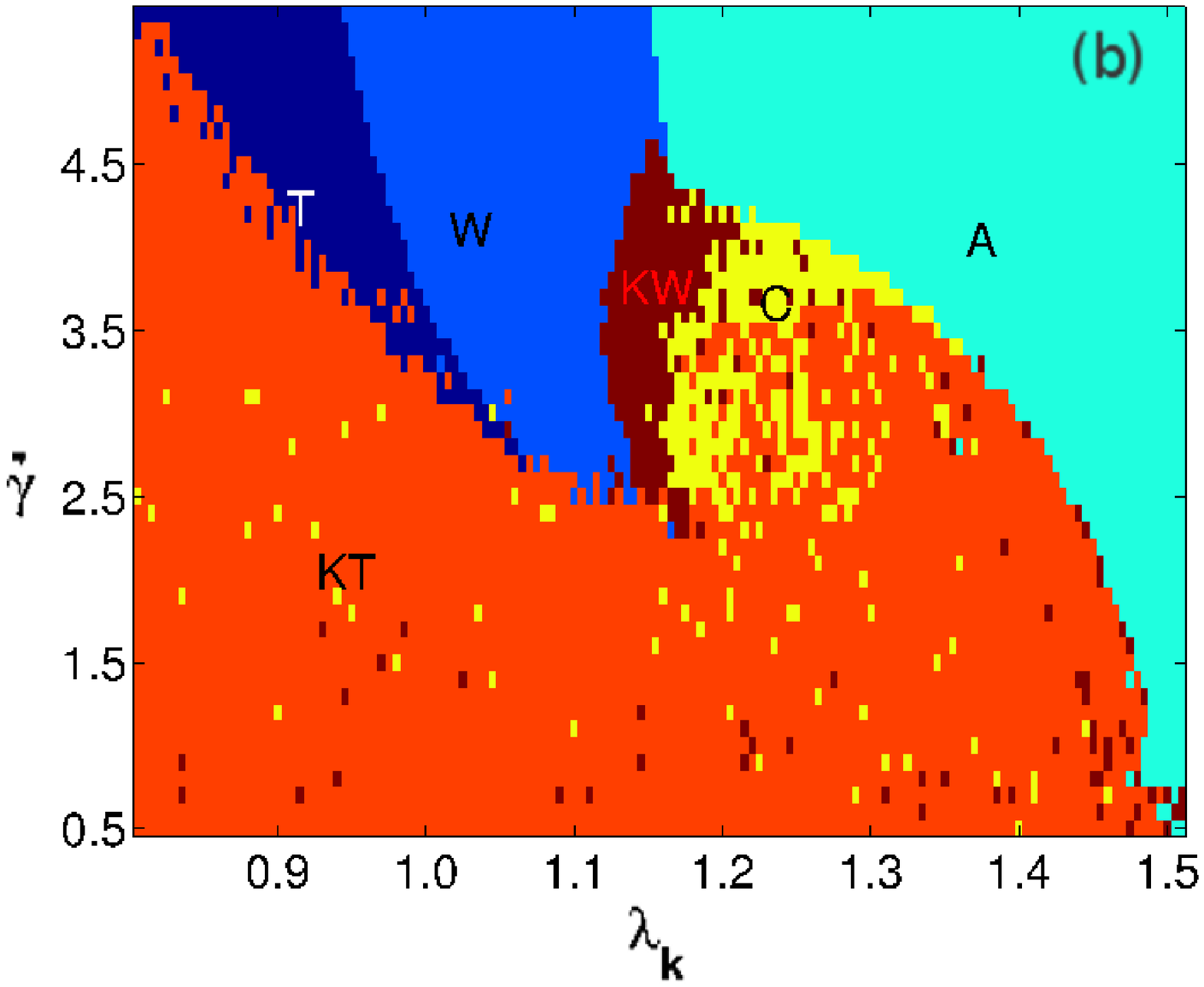}
\caption{[Color Online]
Phase diagram summarizing the local dynamical behaviour of the spatially coupled
one-dimensional system, with $\lambda_k$ plotted on the $x$ axis and 
$\dot{\gamma}$ on the $y$ axis. 
As before,  T denotes the tumbling state, W  the wagging state, KT  the  kayak-tumbling, KW the kayak-wagging state, A the
aligned state and C the state of complex dynamics. The spatial coupling constant
$\epsilon = 0.1$ (a) and 0.5 (b), for a ring of 200 lattice points. The location of the states in the phase
diagram is largely similar to that of Fig.~\ref{local_phase2} with the exception that, at isolated points, mainly
within the KT phase, one sees complex behaviour. The regime in the phase diagram occupied by the C phase
shrinks as the coupling constant $\epsilon$ is increased.}
\label{local_phase3}
\end{center}
\end{figure}
Fig.~(\ref{local_phase3}) shows the dynamical phases  exhibited by a
generic site randomly chosen from the one dimensional ring. The sites
are coupled according to the scheme given in Eq.~\ref{2}, with
coupling constant $\epsilon = 0.1$ (a) and $\epsilon= 0.5$ (b). 

It is evident from comparisons with Fig.~(\ref{local_phase2}) that the local dynamics of a 
generic site in the coupled system is  similar to the uncoupled case. This indicates that
spatial coupling does not  alter the  nature of the  local dynamics qualitatively.
The most significant influence of spatial coupling occurs near the C region, which appear to 
be somewhat broadened with spatial coupling, while
the coexistence regimes are reduced in size. 
In addition, the fairly uniform KT state is now ``studded'' with points displaying complex behaviour. 
This indicates coexistence of complex and KT behaviour, with certain initial states leading to complex 
dynamics, while others lead to a uniform KT state. (It is difficult to determine whether the complex behaviour 
we see is a very long transient or true asymptotic behaviour.)
The tumbling T and wagging W
regions, however, are very stable. 

\subsubsection{Local behaviour of regular regions}

\begin{figure}[!h]
\begin{center}
\includegraphics[width=4.0in]{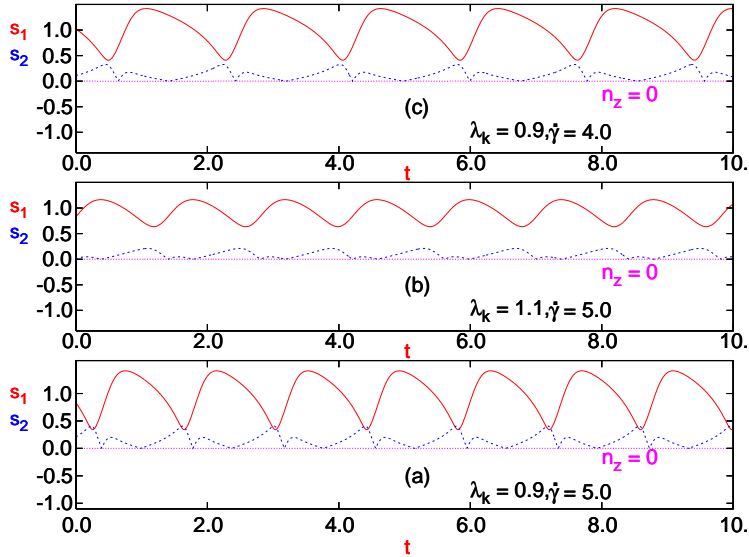}
\caption{[Color Online] Local dynamics in a ring of 200 lattice points, with
  coupling $\epsilon=0.1$, showing the temporal evolution of $s_1$,
  $s_2$ and $n_z$. These are displayed for (a) $s_1$ ,$s_2$ and $n_z$  with
   (a) $\lambda_k = 0.9$ and $\dot{\gamma} = 5.0$ (b) $\lambda_k = 1.1$ and
  $\dot{\gamma} = 5.0$, and (c) $\lambda_k = 0.9$ and $\dot{\gamma} =
  4.0$. These states are all drawn from the T and W parts of the local phase diagram. 
  Note that $n_z=0$ in all these states whereas $s_1$ and $s_2$ are periodic functions
  of time.}
\label{oned_single_siteg5}
\end{center}
\end{figure}

\begin{figure}[!h]
\begin{center}
\includegraphics[width=4.0in]{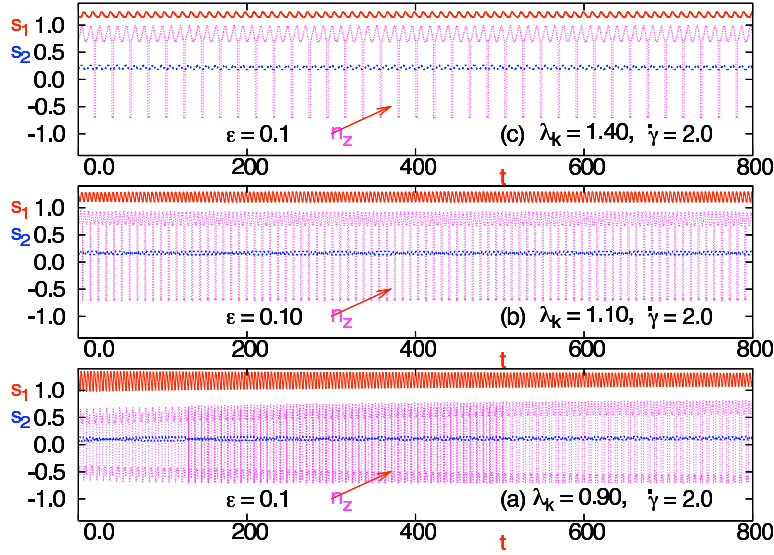}
\caption{[Color Online] Local dynamics in a ring of 200 lattice points, with
  coupling $\epsilon=0.1$, showing the temporal evolution of $s_1$,
  $s_2$ and $n_z$. These are displayed for for (a) $\lambda_k = 0.9$ and $\dot{\gamma} = 2.0$
  (b) $\lambda_k = 1.1$ and $\dot{\gamma} = 2.0$, and (c) $\lambda_k =
  1.4$ and $\dot{\gamma} = 2.0$. These points are drawn from the KT and KW part of the
  local phase diagram and represent states in which the director exhibits out-of-plane fluctuations
  {\it i.e.} $n_z \neq 0$. However, $s_1$ and $s_2$ continue to exhibit regular, periodic oscillations.}
\label{oned_single_siteg2}
\end{center}
\end{figure}

Figs.~(\ref{oned_single_siteg5})-(\ref{oned_single_siteg2}) show
the value of the scalar order parameter $s_1$, the biaxiality parameter
$s_2$ and the $z$ component of the director ${\bf n}$. 
Fig.~(\ref{oned_single_siteg5})  is obtained using parameter values appropriate
to the T and W regions of the local phase diagram,
with a coupling constant $\epsilon = 0.1$.  This displays completely
regular behaviour, with these quantities varying periodically keeping the director 
in the vorticity plane. 
Fig.~(\ref{oned_single_siteg2})  is obtained using parameter values appropriate
to the KT and KW regions of the local phase diagram and indicate that the director
can now fluctuate out of plane whereas all quantities vary smoothly and periodically.

\begin{figure}[!h]
\begin{center}
\includegraphics[width=4.0in]{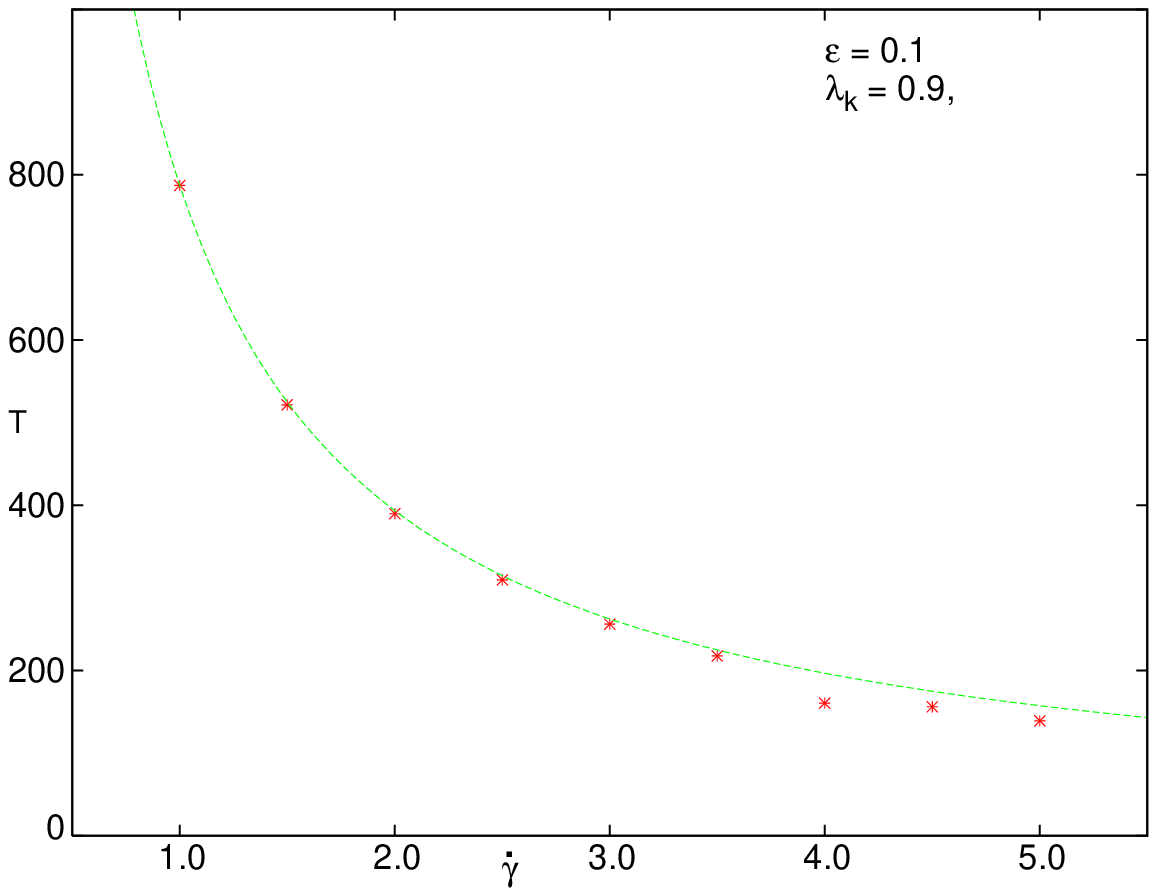}
\caption{[Color Online] Time Period (T) vs. $\dot{\gamma}$, for the case of a ring
  of 200 lattice points, with coupling $\epsilon=0.1$ and with $\dot{\gamma}$ varying
  across the T and KT regions of the phase diagram. The numerical
  results are represented as points and the fitted curve is: $T \sim
  \frac{1}{\dot{\gamma}}$.}
\label{oned_time_period}
\end{center}
\end{figure}

Figure~(\ref{oned_time_period}) shows the local time period with which these
quantities fluctuate as a function of $\dot{\gamma}$ at $\lambda_k = 0.9$.
The time period is inversely
proportional to shear rate, as is evident from the fit to the data points. 
\subsubsection{Local Dynamics in the Complex Region}
\begin{figure}
\begin{center}
\includegraphics[width=4.0in]{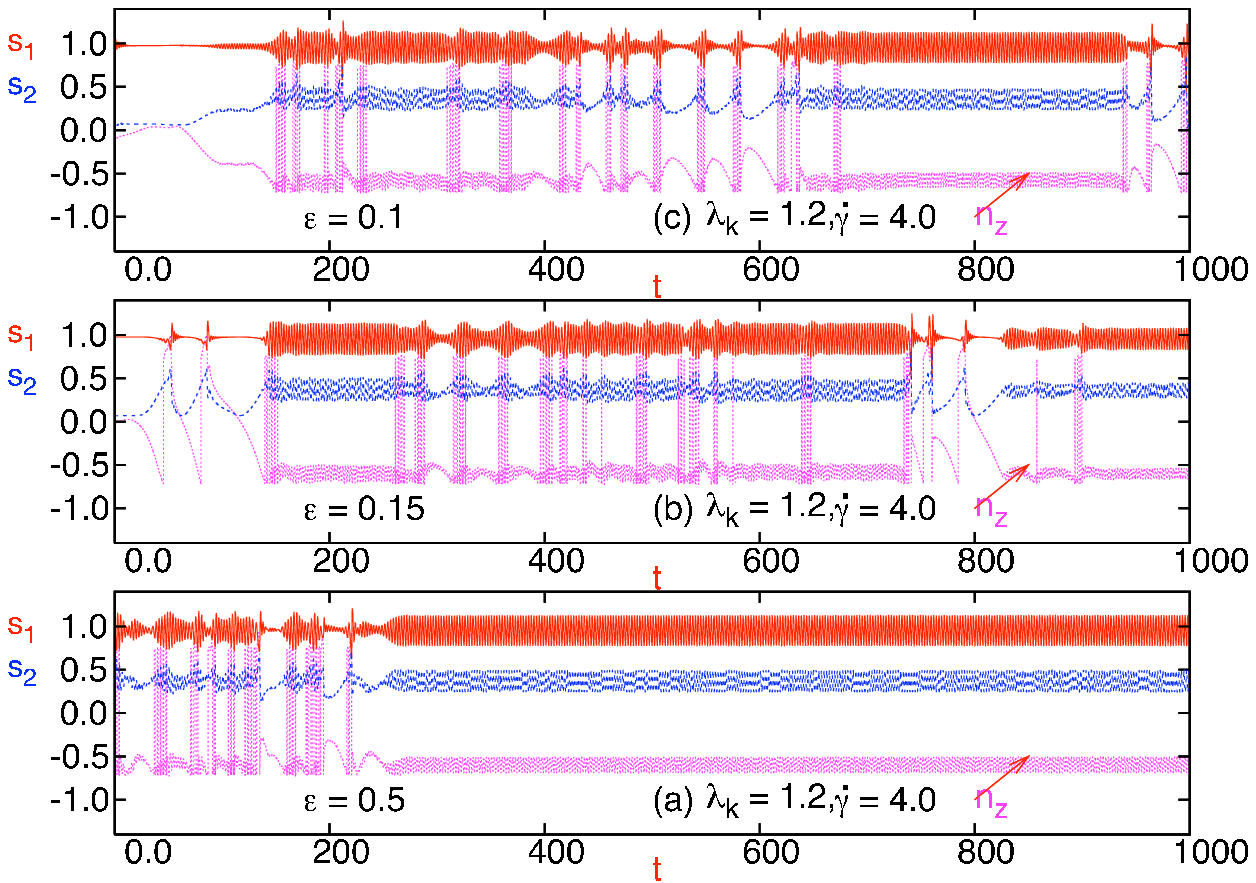}
\caption{[Color Online] Local dynamics of a ring of 200 lattice points, with
  $\lambda_k = 1.2$ and $\dot{\gamma} = 4.0$, showing the time
  evolution of $s_1$, $s_2$ and $n_z$. These are shown for (a) $\epsilon = 0.5$ (b)
 $\epsilon = 0.15$ and (c) $\epsilon = 0.1$, illustrating behaviour in the
 complex or C regime. Note that regular time-periodic behaviour is favoured at
 large values of the spatial coupling constant $\epsilon$, following an initial transient.}
\label{oned_single_siteg4}
\end{center}
\end{figure}

The local behaviour in the complex region, 
denoted by $C$ in the local phase
diagram, is exhibited in Fig.~(\ref{oned_single_siteg4}), which
shows $s_1$ and $s_2$ and $n_z$. The results suggest that the sites
display {\em intermittent} behaviour. These results are obtained for
parameter values at 
the boundary of the complex region and the kayak-wagging region, with
paramters $\dot{\gamma} = 4.0$ and $\lambda_k = 1.2$. In part (a), the
coupling constant $\epsilon$=  $0.5$, in part (b) $\epsilon = 0.15$
and in part (c) $\epsilon = 0.1$. All  of these show qualitatively similar temporally
intermittent behaviour.
\begin{figure}
\begin{center}
\includegraphics[width=4.0in]{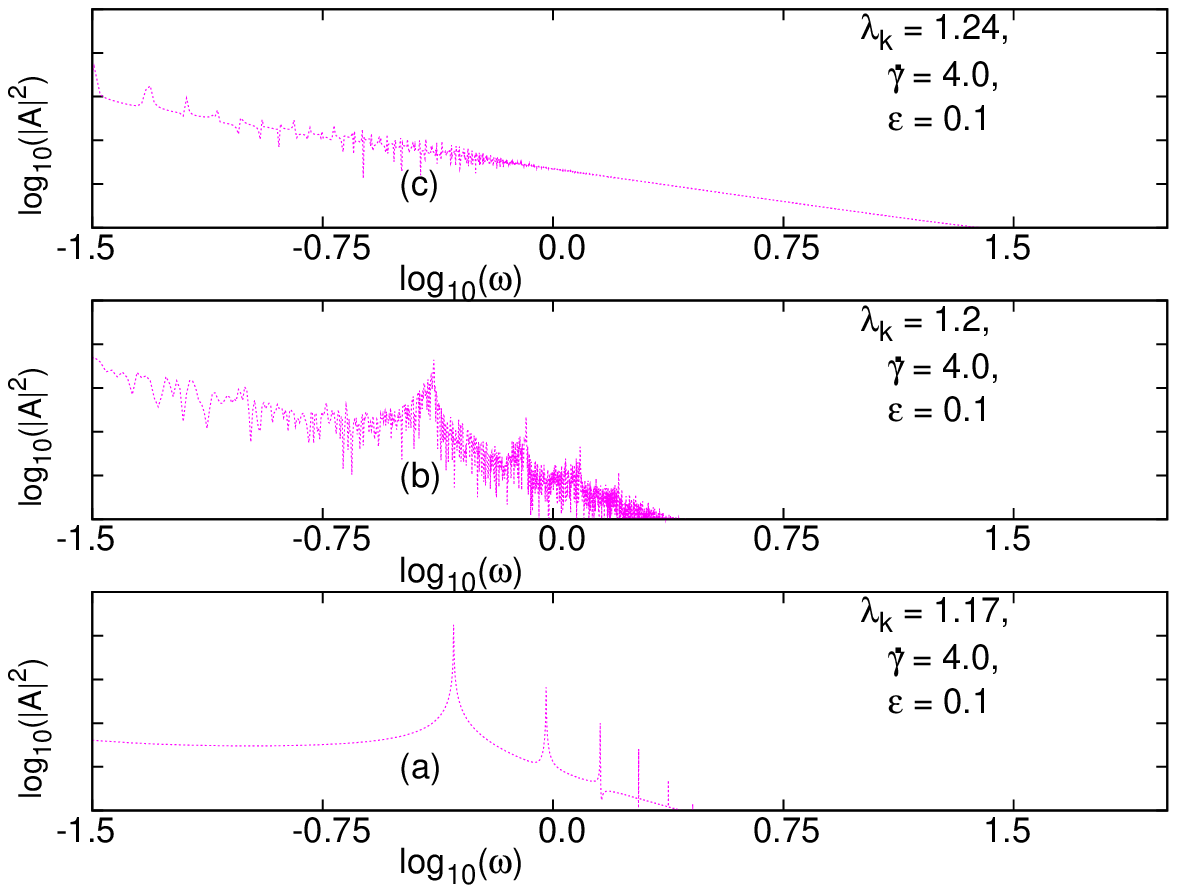}
\caption{[Color Online] Log-Log plot of the Fourier transform vs. frequency, for (a)
  $\lambda_k = 1.17$, (b) $\lambda_k = 1.20$ and (c) $\lambda_k =
  1.24$. Here $\epsilon = 0.1$ and $\dot{\gamma} = 4.0$. The
  lattice is a ring of $200$ sites. Note that for $\lambda_k = 1.17$ (the wagging region), the regular 
  oscillations show up as a delta function in the fourier transform. In the C or complex region,
  a smooth distribution of frequencies is seen, with a $1/f^2$ falloff.
  } 
\label{oned_single_site_ft_g4}
\end{center}
\end{figure}
The fourier transform of the time series of stress calculated at a generic site  and
plotted on a doubly 
logarithmic scale is shown in Fig.~(\ref{oned_single_site_ft_g4}).  The
spectrum in (c) of Fig.~(\ref{oned_single_site_ft_g4}) fits the
relation $P(f) \sim 1/f^2$.

\subsection{Spatio-temporal coherence and dynamics}
In order to quantify the degree of spatial coherence, we calculate the
following quantity for the one-dimensional lattice:
\begin{equation}
\overline{d} = \sqrt{\frac{1}{N T}\sum_{t = 1}^T\sum_{i = 1}^N (a_0^t(i) - \overline{a}^t)^2},
\label{oned_dav}
\end{equation}
where
\begin{equation}
\overline{a^t} = \frac{1}{N}\sum_{i = 1}^N a^t_0(i).
\label{oned_av}
\end{equation}
We have calculate such a spatial coherence parameter for one specific
component of the vector $(a_0, \ldots, a_4)$; however, qualitatively
similar results are obtained for other components as well as for the
full local stress, in the C region. When $\overline{d}$ tends to zero the degree of synchronization of the
local variables is very high. On the other hand large $\overline{d}$
indicates low spatial synchronization, arising from a wide
distribution of values of the local variables in the lattice.  This
quantity thus serves as a global order parameter characterizing the
smoothness of the spatial patterns exhibited by the evolution of the map.

\begin{figure}
\begin{center}
\includegraphics[width=2.95in]{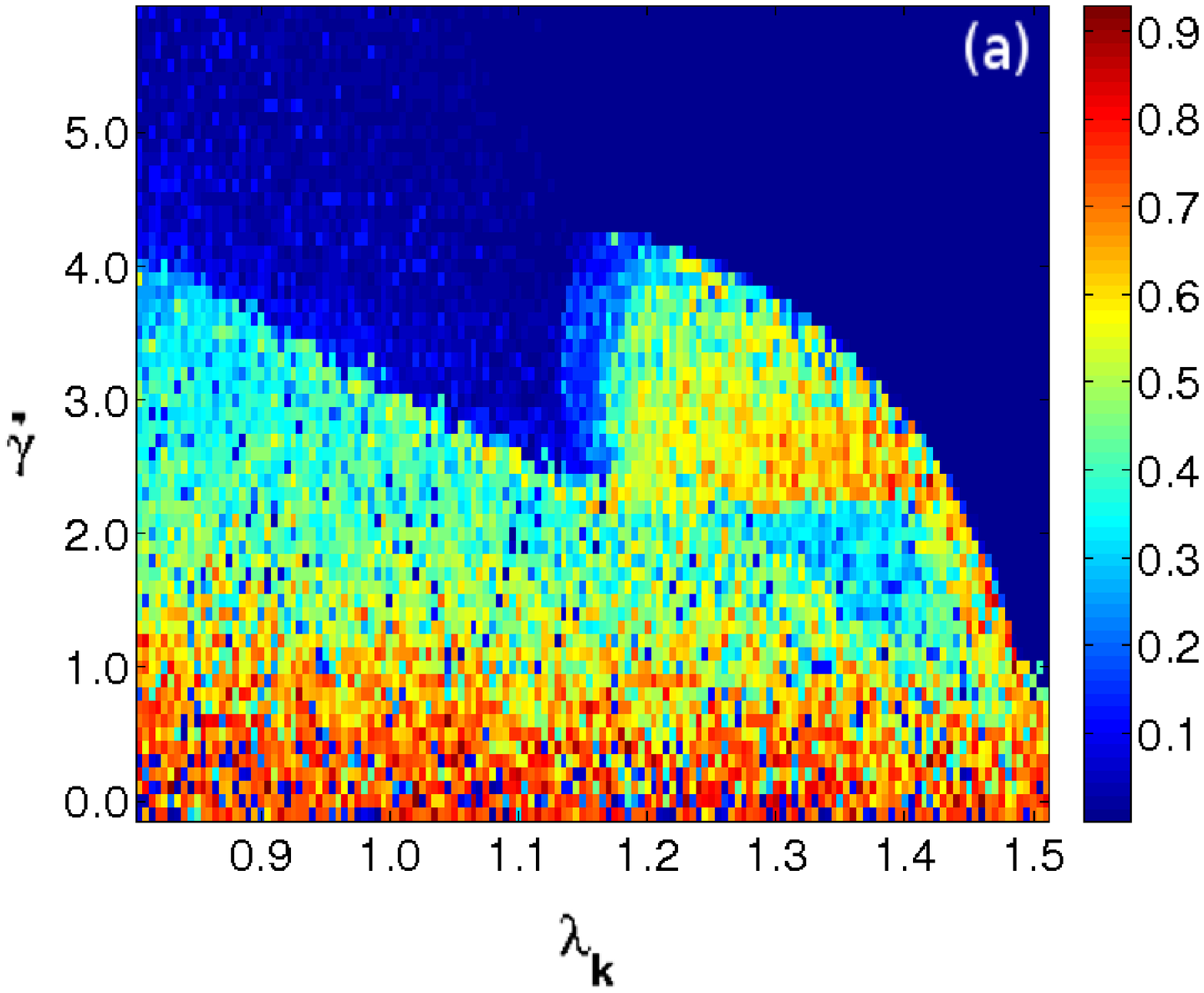}
\includegraphics[width=3.0in]{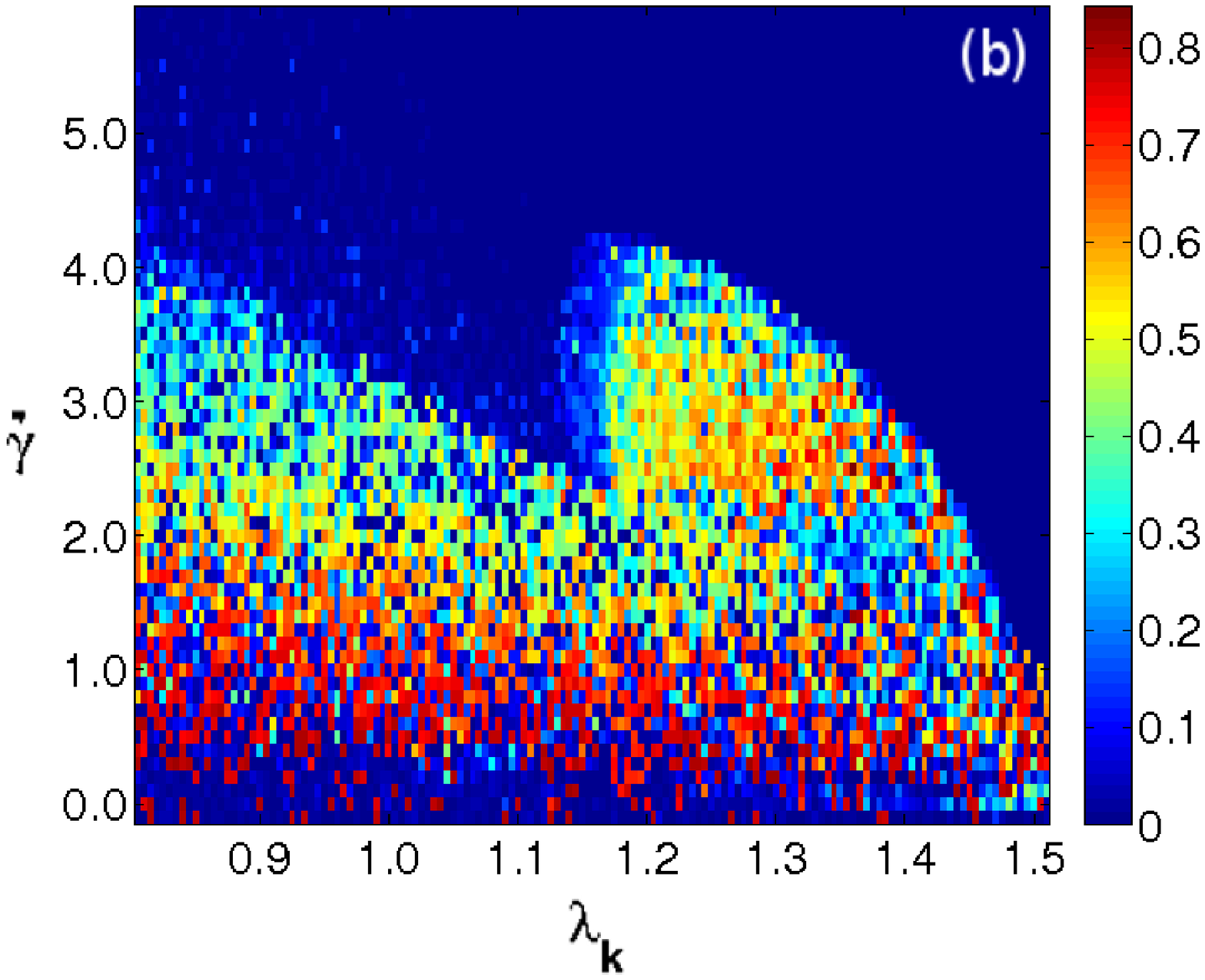}
\caption{[Color Online] Average deviation from the mean value for $\epsilon = 0.1$
  (a) and $\epsilon = 0.5$ (b), with $\lambda_k$ on the
  $x$-axis and $\dot{\gamma}$ on the $y$ axis. Note that large fluctuations (roughness)
  are seen in the KW and C regions. These data are for the 1-d system wiith the number of
  sites $N=200$ and parameters as indicated on the figure.}
\label{oned_av_dev}
\end{center}
\end{figure}

Fig.~(\ref{oned_av_dev}) shows the time average of the deviation
$<\overline{d}>$ of $a_0$ from the average value $a_0$. To compute this,
we first calculate the instantaneous deviation $\overline{d}$ via
Eqs.~\eqref{oned_av} and \eqref{oned_dav}, and then find the
long-time average of this quantity. The spatial profile of the regular
region with low $<\overline{d}>$ is characterised  either by
spatiotemporal fixed behaviour with all sites aligned, or spatial
uniformity and temporal periodicity. There are also cases in the
regular region where the sites, though not completely synchronized in
space, are nevertheless phase synchronized.

\subsubsection{Spatio-temporal dynamics in the regular region}
\begin{figure}
\begin{center}
\includegraphics[width=3.0in]{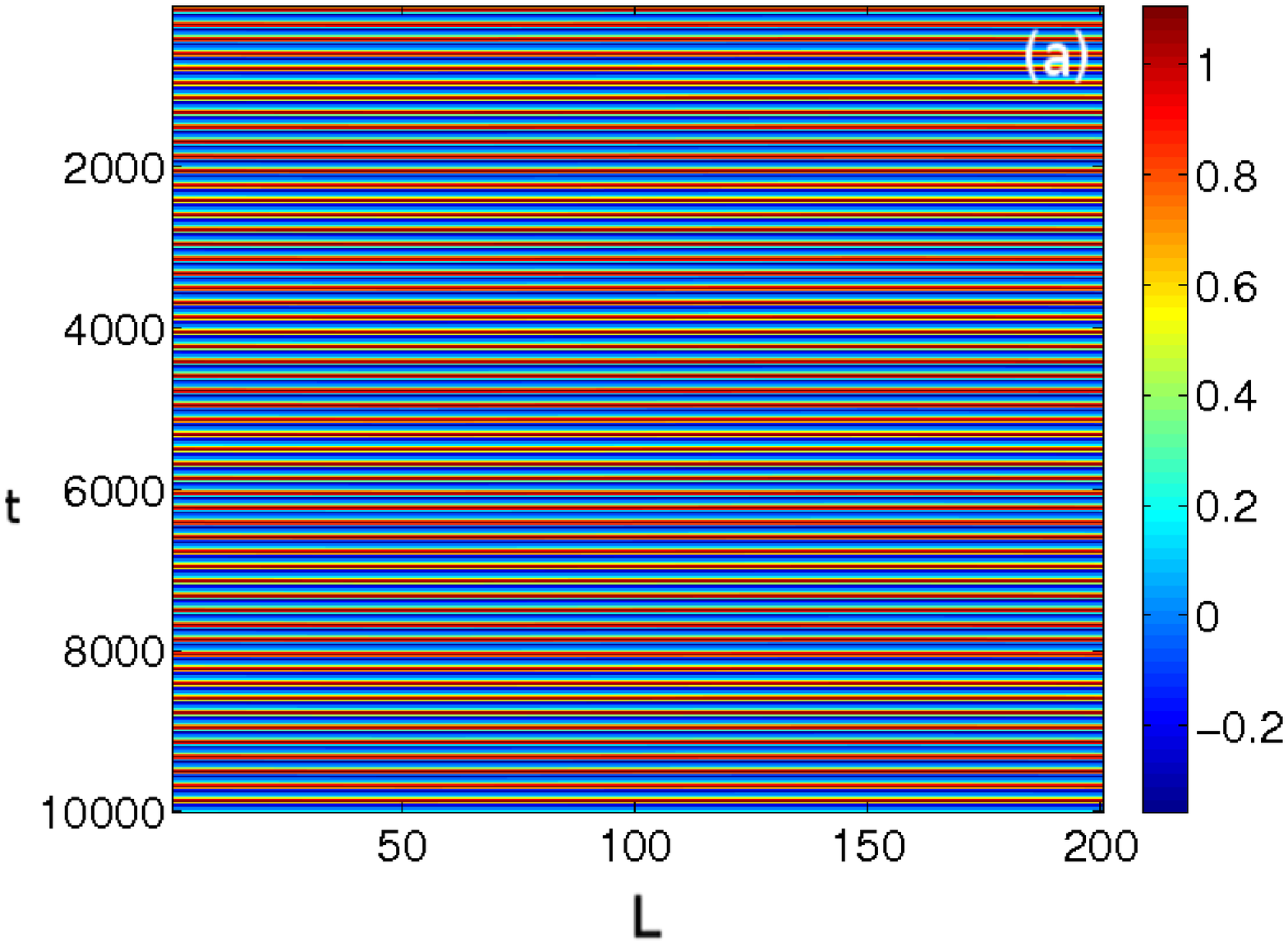}
\includegraphics[width=3.0in]{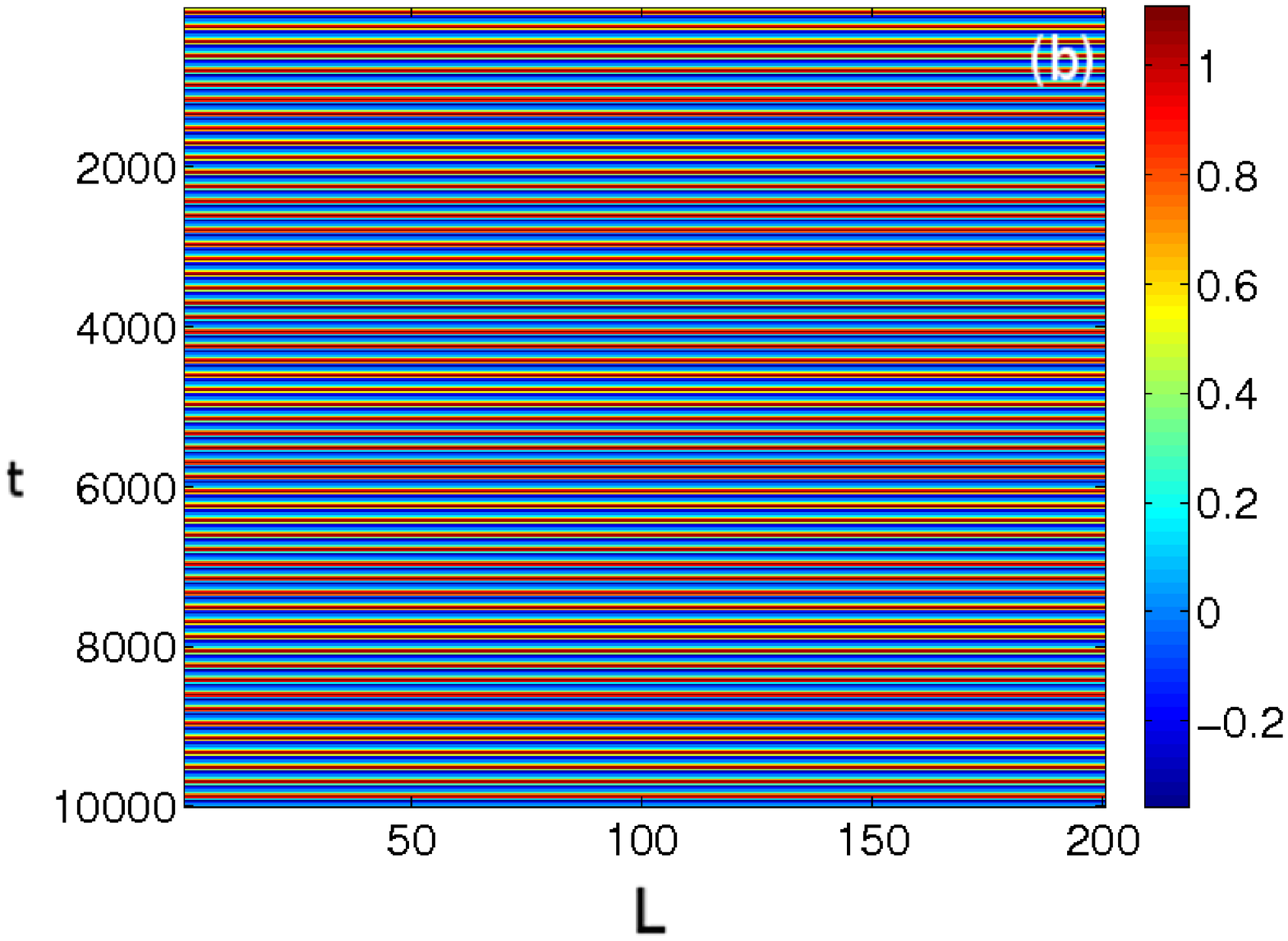}
\caption{[Color Online] Density plot of order parameter part of shear stress. Here
  $\lambda_k$ = 1.1, $\dot{\gamma}$ =4.0 and $\epsilon$ = 0.1 (a)
  and 0.5 (b). Time steps (running from top to bottom) are on the
  $y$-axis, and the lattice site index ($i=1, 200$) is on the
  $x$-axis. These figures represent space-uniform and time-periodic states, obtained using 
  parameter values corresponding to the T region of the phase diagram.}
\label{ondstg4l1_1}
\end{center}
\end{figure}

Fig.~(\ref{ondstg4l1_1}) displays the space-time plot for $\dot{\gamma} =
4$, $\lambda_k = 1.1$ and coupling constant $\epsilon = 0.1$ (a)
and $0.5$ (b). The $x$-axis displays the lattice index and time is  shown on
$y$-axis, increasing from top to bottom. The profile is not spatially uniform and periodic in time
for very weak coupling.  As the coupling is increased, the system acquires
spatial coherence and temporal periodicity.

%==============================================================================================%
%\subsubsection{Space profile of complex region}
%==============================================================================================%
\subsubsection{Spatio-temporal dynamics in the complex region}

\begin{figure}
\begin{center}
\includegraphics[width=2.85in]{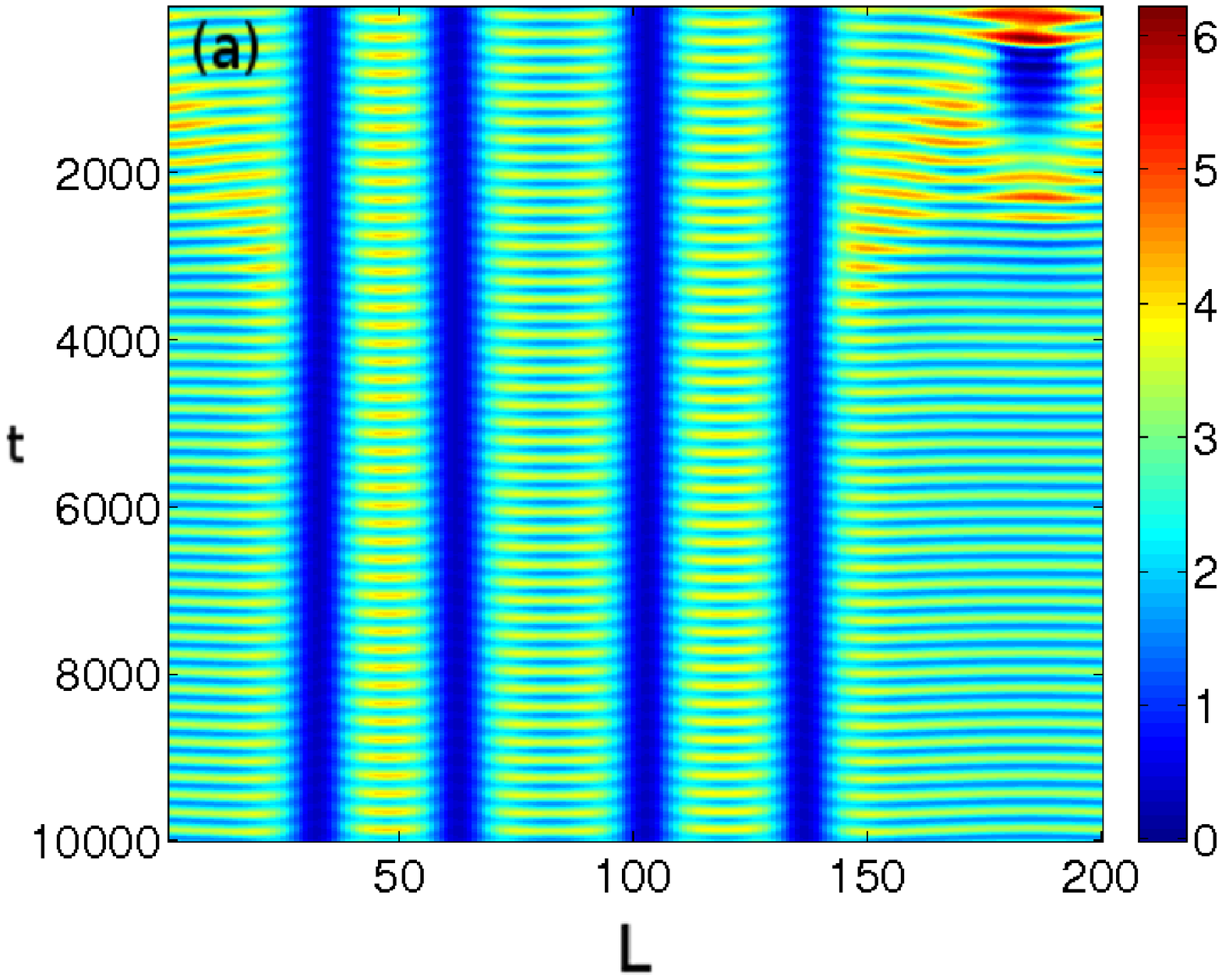}
\includegraphics[width=3.0in]{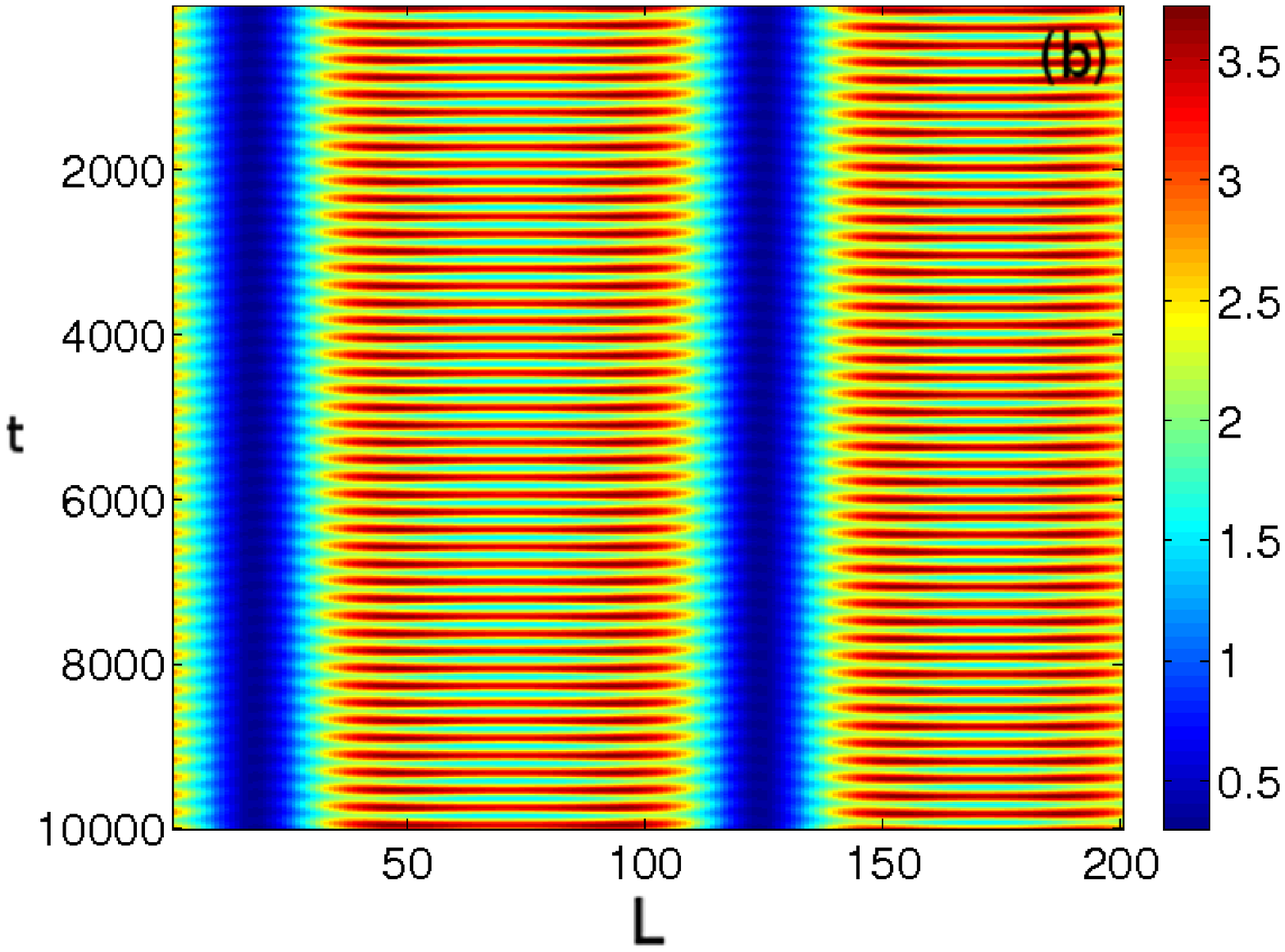}
\caption{[Color Online] Density plot of order parameter part of shear stress. Here
  $\lambda_k$ = 1.17, $\dot{\gamma}$ =4.0 and $\epsilon$ = 0.1 (a)
  and 0.5 (b). Time steps (running from top to bottom) are on the
  $y$-axis, and the lattice site index ($i=1, 200$) is on the
  $x$-axis. These figures represent space non-uniform and time-periodic states, obtained using 
  parameter values corresponding to the KW region of the phase diagram}
\label{ondstg4l1_17}
\end{center}
\end{figure}

\begin{figure}
\begin{center}
\includegraphics[width=2.95in]{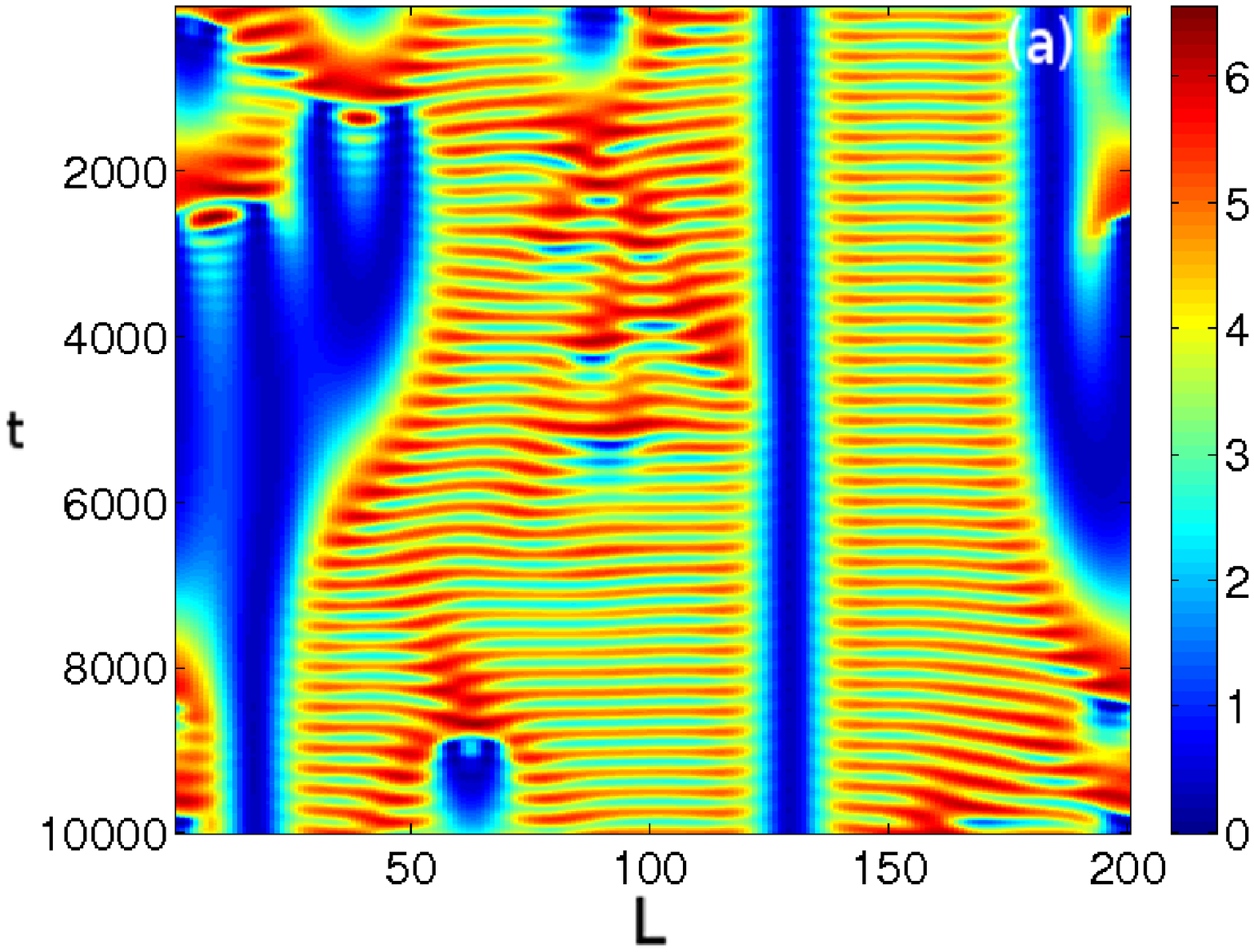}
\includegraphics[width=3.0in]{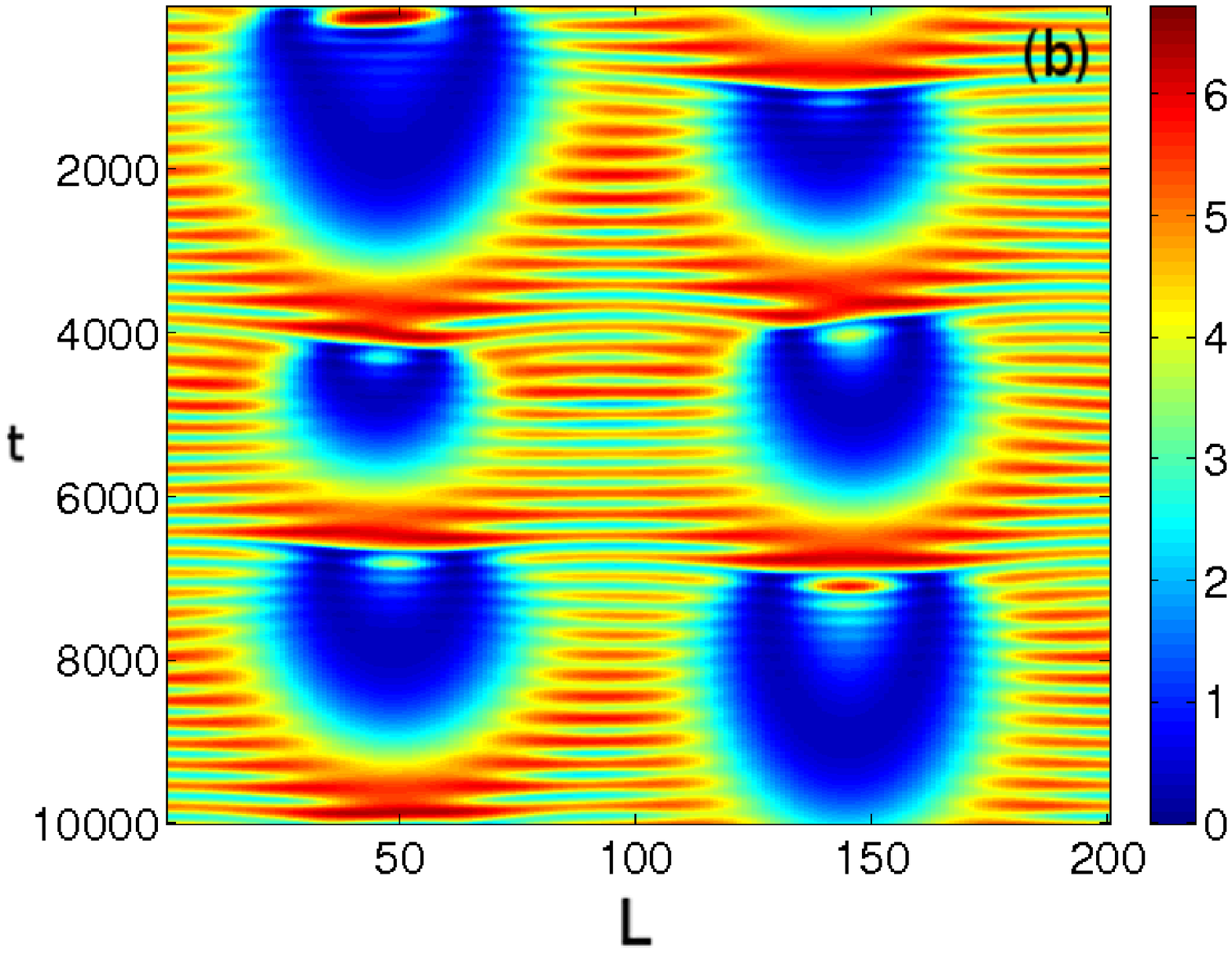}
\caption{[Color Online] Density plot of order parameter part of shear stress. Here
  $\lambda_k$ = 1.20, $\dot{\gamma}$ =4.0 and $\epsilon$ = 0.1 (left)
  and 0.5 (right). Time steps (running from top to bottom) are on the
  $y$-axis, and the lattice site index ($i=1, 200$) is on the
  $x$-axis. 
  These figures illustrate how time-periodic regions are interspersed with domains of 
  fixed point behaviour, reminiscent of spatiotemporal intermittency.
  The parameter values corresponding to the C region of the phase diagram, in a regime
  where the chaos is weak.}
\label{ondstg4l1_20}
\end{center}
\end{figure}

\begin{figure}
\begin{center}
\includegraphics[width=3.0in]{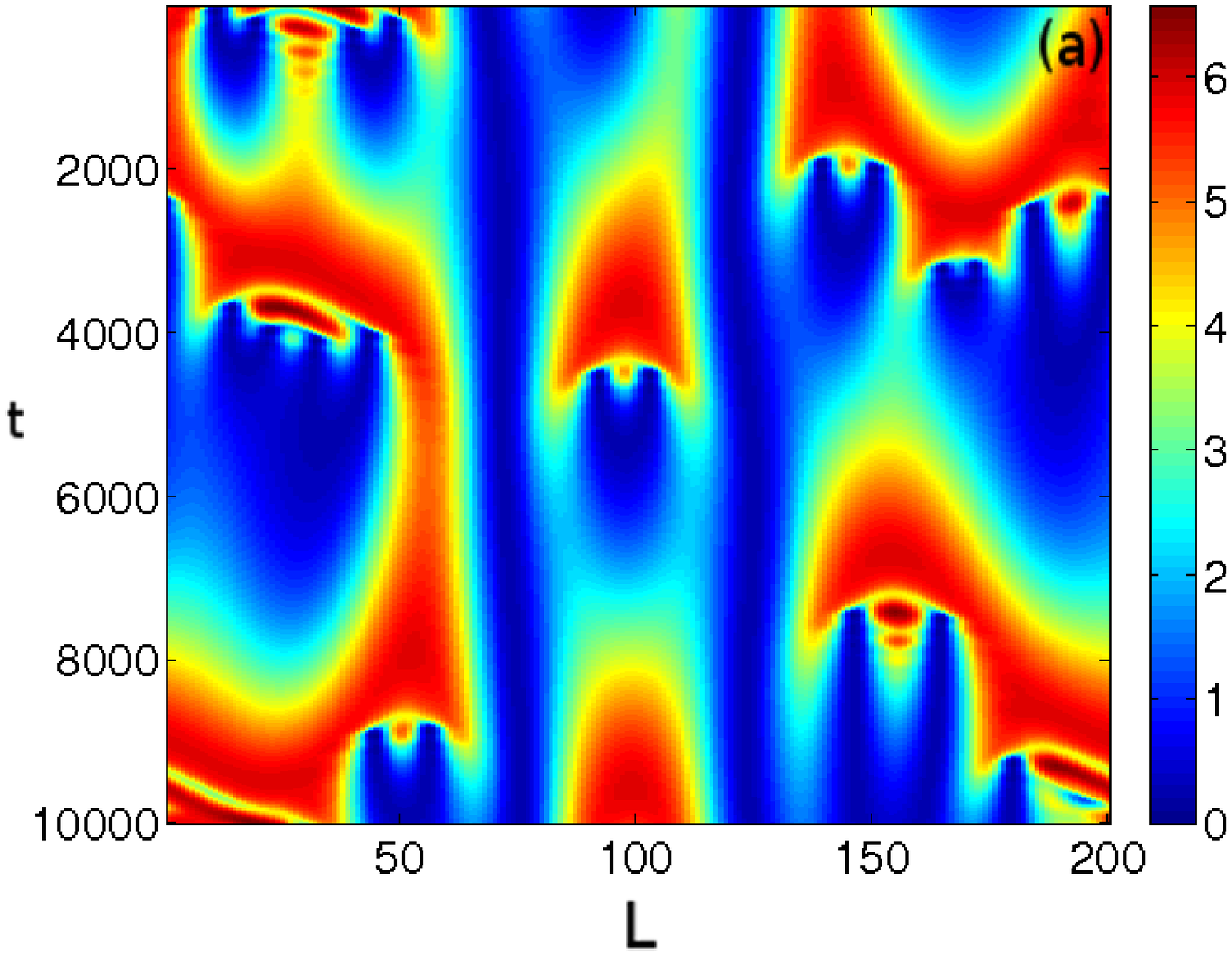}
\includegraphics[width=3.0in]{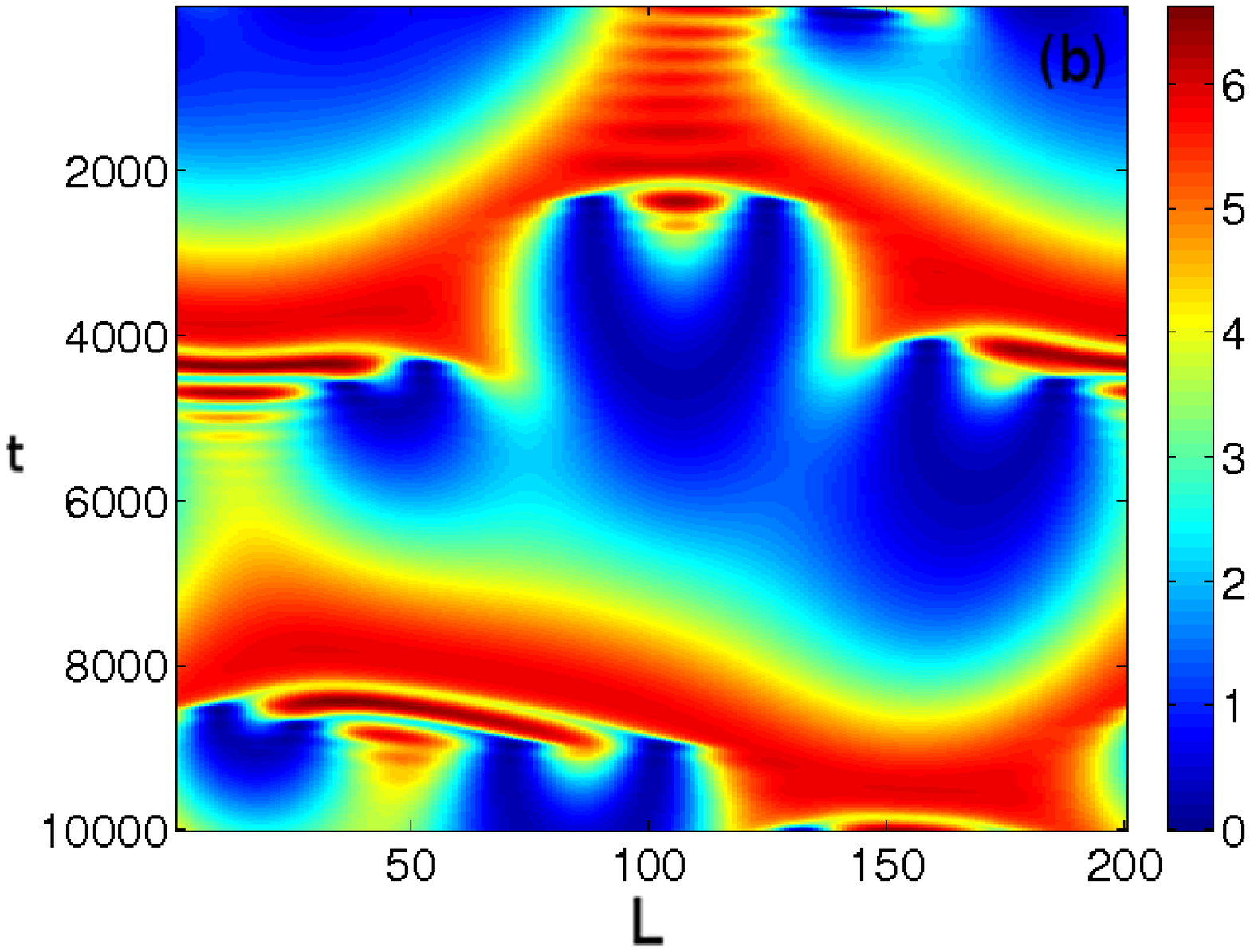}
\caption{[Color Online] Density plot of order parameter part of shear stress. Here
  $\lambda_k$ = 1.24, $\dot{\gamma}$ =4.0 and $\epsilon$ = 0.1 (a)
  and 0.5 (b). Time steps (running from top to bottom) are on the
  $y$-axis, and the lattice site index ($i=1, 200$) is on the
  $x$-axis. These figures illustrate non-uniform, time-varying states characteristic
  of spatio-temporally chaotic behaviour.
  The parameter values correspond to the C region of the phase diagram, in a regime
  where the chaos is strong. Note that larger values of $\epsilon$ lead to larger and more
  uniform spatial structures.}
\label{ondstg4l1_24}
\end{center}
\end{figure}

The spatiotemporal behaviour of a representative case in the C or complex
region is displayed in Fig.~(\ref{ondstg4l1_17}), where $\dot{\gamma} =
4$, $\lambda_k = 1.17$ and coupling constant $\epsilon$ = 0.1 (a)
and 0.5 (b).  It is evident that the space-time profile splits into bands, i.e. clusters of synchronized sites, where the local 
dynamics is either fixed (blue) or time-periodic (stripes).
As we increase $\lambda_k $ (with $\dot{\gamma} = 4$) in Figs.
~(\ref{ondstg4l1_17} - ~\ref{ondstg4l1_24}) the length scale of the
spatio-temporally intermittent pattern increases, finally yielding to the aligned region.
This progression from frozen localized kinks/domains of fixed points in the spatial background of time-periodic behaviour, to infective bursts bearing the signature of spatiotemporal intermittency, is seen in many systems \cite{crutchfieldkaneko}, and often arises from a competition of fixed point patterns and time-periodic and quasi-periodic patterns.

%For the figure (\ref{ondstg3l124)}, we find the
%same changes for $\dot{\gamma} = 3$, as observed for $\dot{\gamma} =
%4$. The system moves from regular states to intermittent states, as we increase $\lambda_k$. 
%So we do not show the representative figures for $\dot{\gamma}$ In the intermittent system there are spatial
%domains of oscillatory behavior, displaying the space-time triangular
%shape characteristic of spatiotemporal intermittency.
 
%==============================================================================================%
%          Figures Related to Space Profile of One dimensional Lattice                         %

%==============================================================================================%

\begin{figure}
\begin{center}
\includegraphics[width=3.0in]{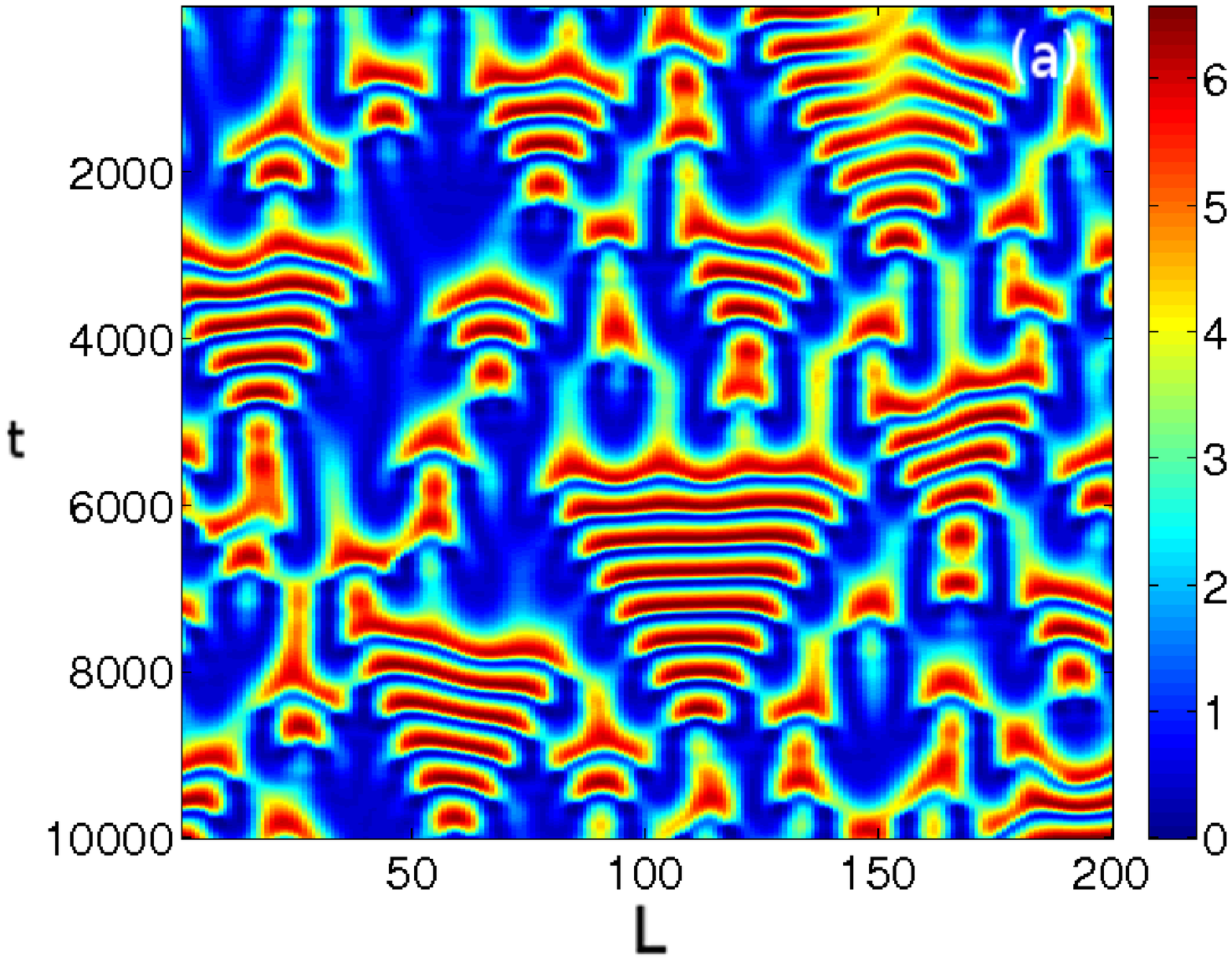}
\includegraphics[width=3.1in]{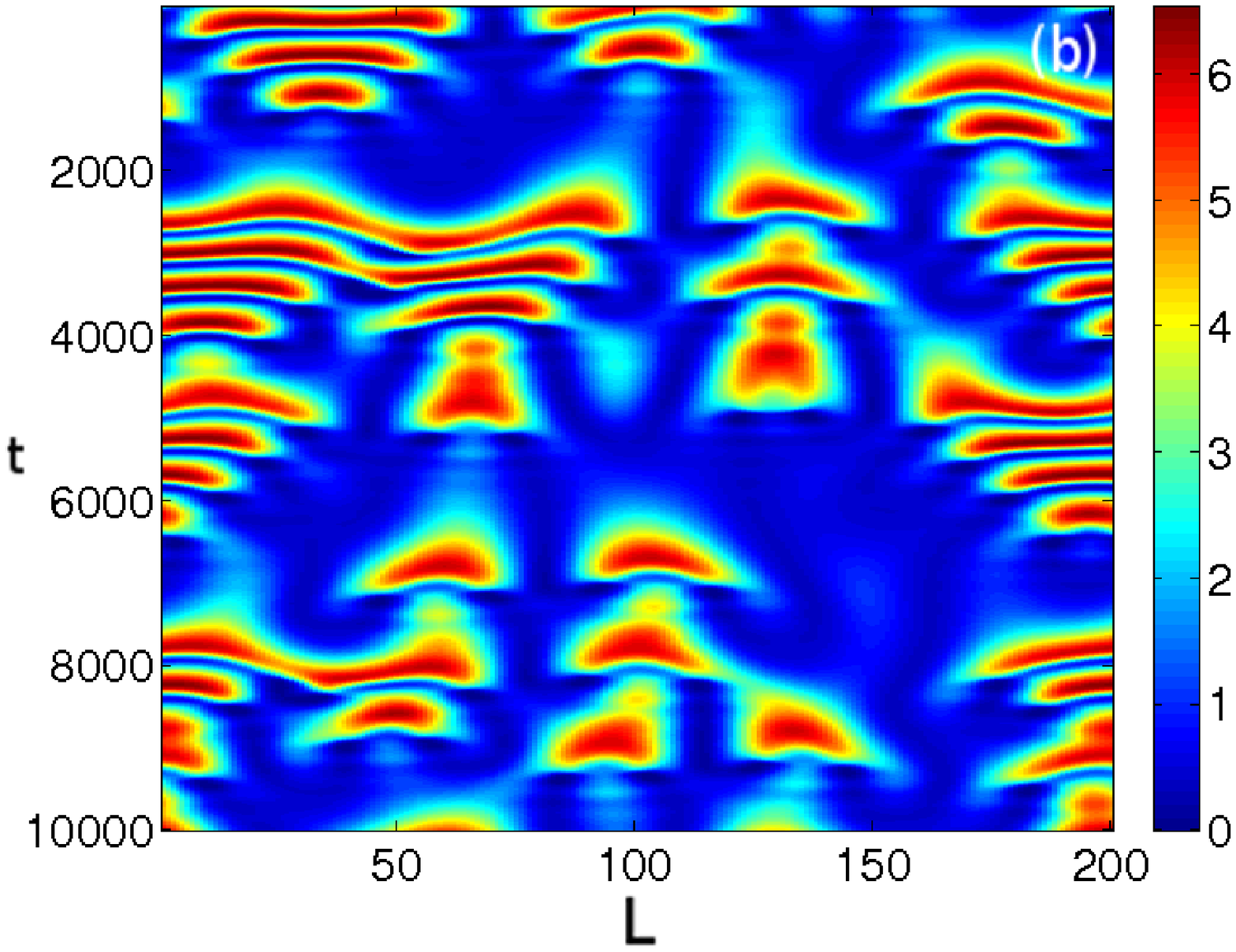}
\caption{[Color Online] Density plot of order parameter part of shear stress. Here
  $\lambda_k$ = 1.24, $\dot{\gamma}$ =3.0 and $\epsilon$ = 0.1 (a)
  and 0.5 (b). Time steps (running from top to bottom) are on the
  $y$-axis, and the lattice site index ($i=1, 200$) is on the
  $x$-axis.}
\label{ondstg3l124}
\end{center}
\end{figure}
%\end{figure}

\begin{comment}
\begin{figure}
\begin{center}
\includegraphics[width=4.0in]{four_trans_g4_l.eps}
\caption{[Color Online] Fourier transform of the stress, averaged over initial
  conditions and space, for (a) $\lambda_k = 1.17$ and $\dot{\gamma} =
  4.0$ (b) $\lambda_k = 1.20$ and $\dot{\gamma} = 4.0$ (c) $\lambda_k
  = 1.24$ and $\dot{\gamma} = 4.0$. Here $\epsilon = 0.1$, and lattice
  size is $200$.}
\label{ft}
\end{center}
\end{figure}
\end{comment}

\section{The Two-dimensional Coupled Map Lattice}

%==============================================================================================%
\subsection{Local temporal behaviour}
%==============================================================================================%
We have also explored the dynamics on a two-dimensional
lattice. In the regular regions of the phase diagram, corresponding to the
T,W, KT and KW states, the temporal behavior is very similar to that
of the one dimensional case and is thus not shown separately. We thus 
concentrate on behavior in the complex or C region. 

\begin{figure}
\begin{center}
\includegraphics[width=5.0in]{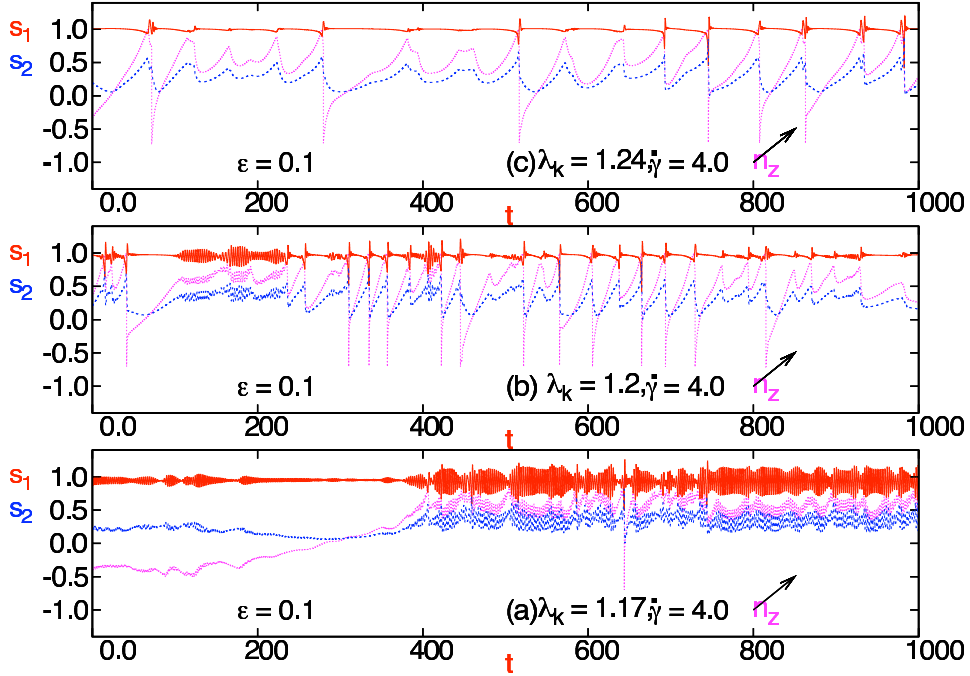}
\caption{[Color Online] Temporal evolution of $s_1$, $s_2$ and $n_z$ in a two
  dimensional lattice of size $102\times102$, with $\epsilon=0.1$, for
  (a) $\lambda_k = 1.17$ and $\dot{\gamma} = 4.0$ (b) $\lambda_k = 1.20$
  and $\dot{\gamma} = 4.0$ (c) $\lambda_k = 1.24$ and $\dot{\gamma} =
  4.0$. All these state points are drawn from the C region of the local phase diagram.
  Note the existence of temporally intermittent behavior, analysed in terms of its
  frequency spectrum in Fig.~\ref{twod_single_site_ft_g4}. }
\label{single_site_g4_2d}
\end{center}
\end{figure}

\begin{figure}
\begin{center}
\includegraphics[width=4.0in]{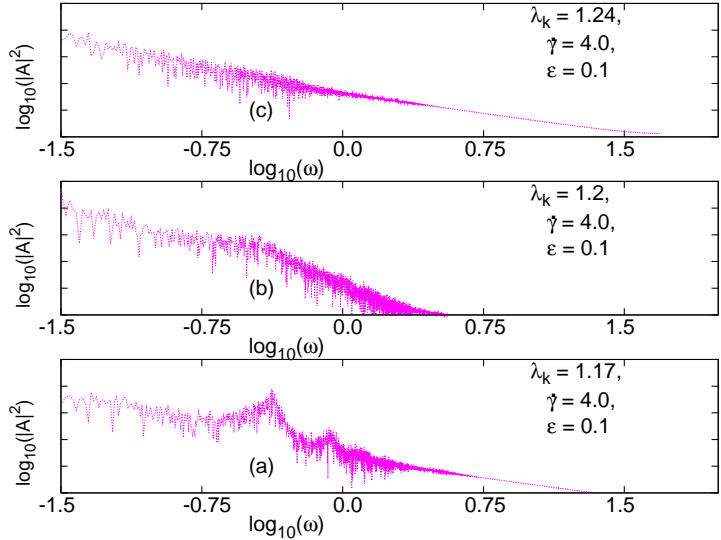}
\caption{[Color Online]  Log-Log plot of the absolute value of the Fourier transform of the stress 
  vs. frequency $\omega$, for (a)   $\lambda_k = 1.17$, (b) $\lambda_k = 1.20$ and (c) $\lambda_k =
  1.24$. Here $\epsilon = 0.1$ and $\dot{\gamma} = 4.0$. The
  lattice contains $100 \times 100$ sites. Note the relatively smooth background, indicating the presence
  of a continuous set of frequencies. The fall-off is consistent with a $1/\omega^2$ behaviour.}
\label{twod_single_site_ft_g4}
\end{center}
\end{figure}

Representative data showing the local temporal dynamics of the
complex region is given in Fig~(\ref{single_site_g4_2d}).
% and part (b) of the figure~(\ref{single_site_g3_2d}).
They show chaotic
behaviour, and there appears to be no qualitative difference between
the one dimensional and two dimensional lattice cases. As in the
one-dimensional lattice, increased coupling strengths suppress the
chaotic region. The log-log plot of the fourier transform is shown in
Fig.~(\ref{twod_single_site_ft_g4}); a similar fit to $P(f) \sim f^{-2}$
of the frequency spectrum of the stress can be obtained, as in the one-dimensional
case.

\subsection{Spatio-temporal behaviour}
To quantify the degree of spatial coherence in
$2$-dimensional lattices we calculate the quantity, generalizing from the
one-dimensional case studied in an earlier section:
\begin{equation}
\overline{d} = \sqrt{\frac{1}{N_1 N_2 T}\sum_{t = 1}^T\sum_{i,j = 1}^{N1,N2} (a_0^t(i,j) - \overline{a}^t)^2},
\label{twod_dav}
\end{equation}
with
\begin{equation}
\overline{a} = \frac{1}{N_1 N_2}\sum_{i,j =1}^{N_1,N_2} a_0(i,j).
\label{twod_av}
\end{equation}
Again, as in the $1$-dimensional case, when $\overline{d}$ tends to
zero the degree of synchronization of the local variables is very
high. On the other hand large $\overline{d}$ indicates low spatial
synchronization, and arises from a wide distribution of values of the
local variables in the lattice. 

\begin{figure}
\begin{center}
\includegraphics[width=3.0in]{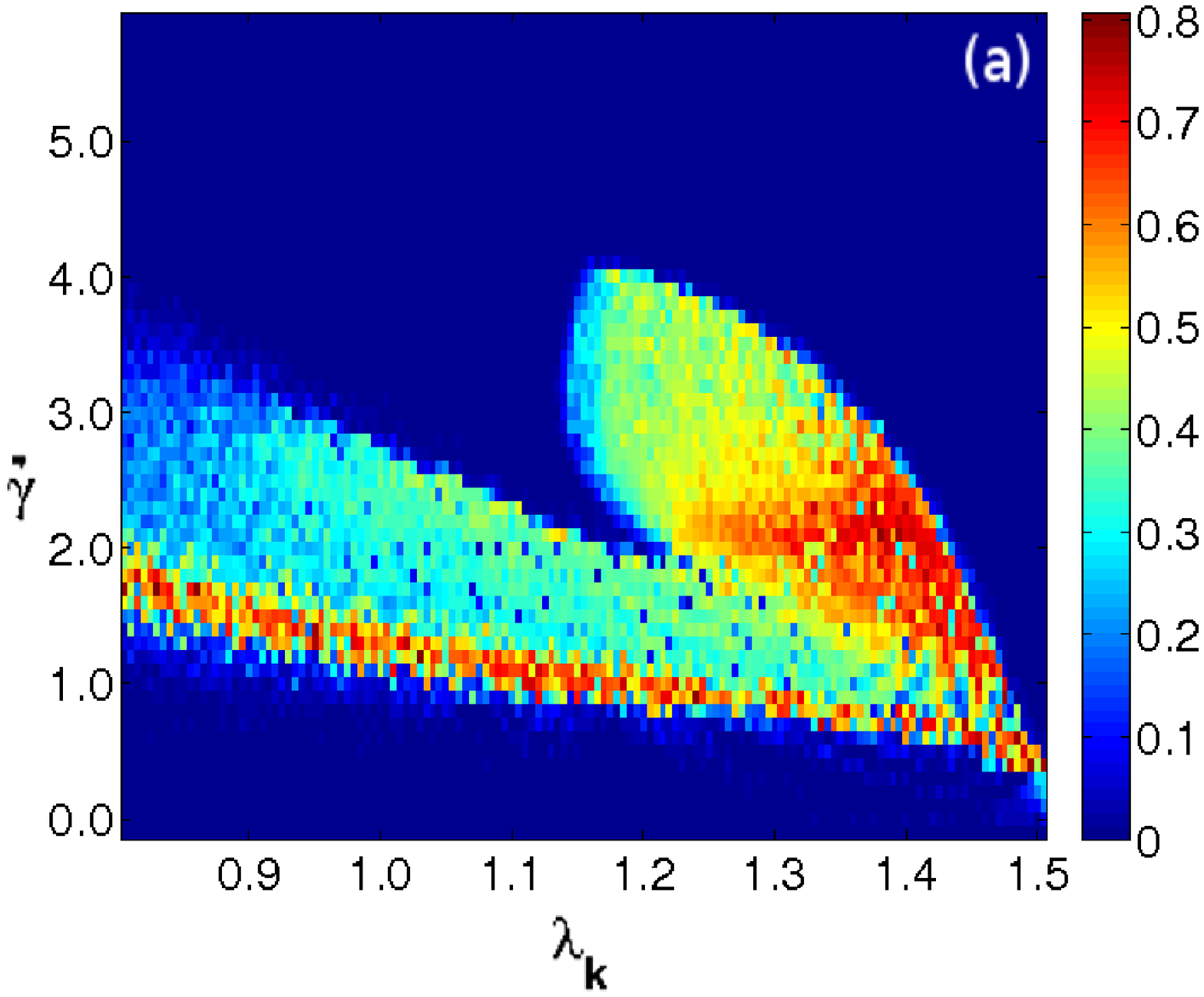}
\includegraphics[width=3.0in]{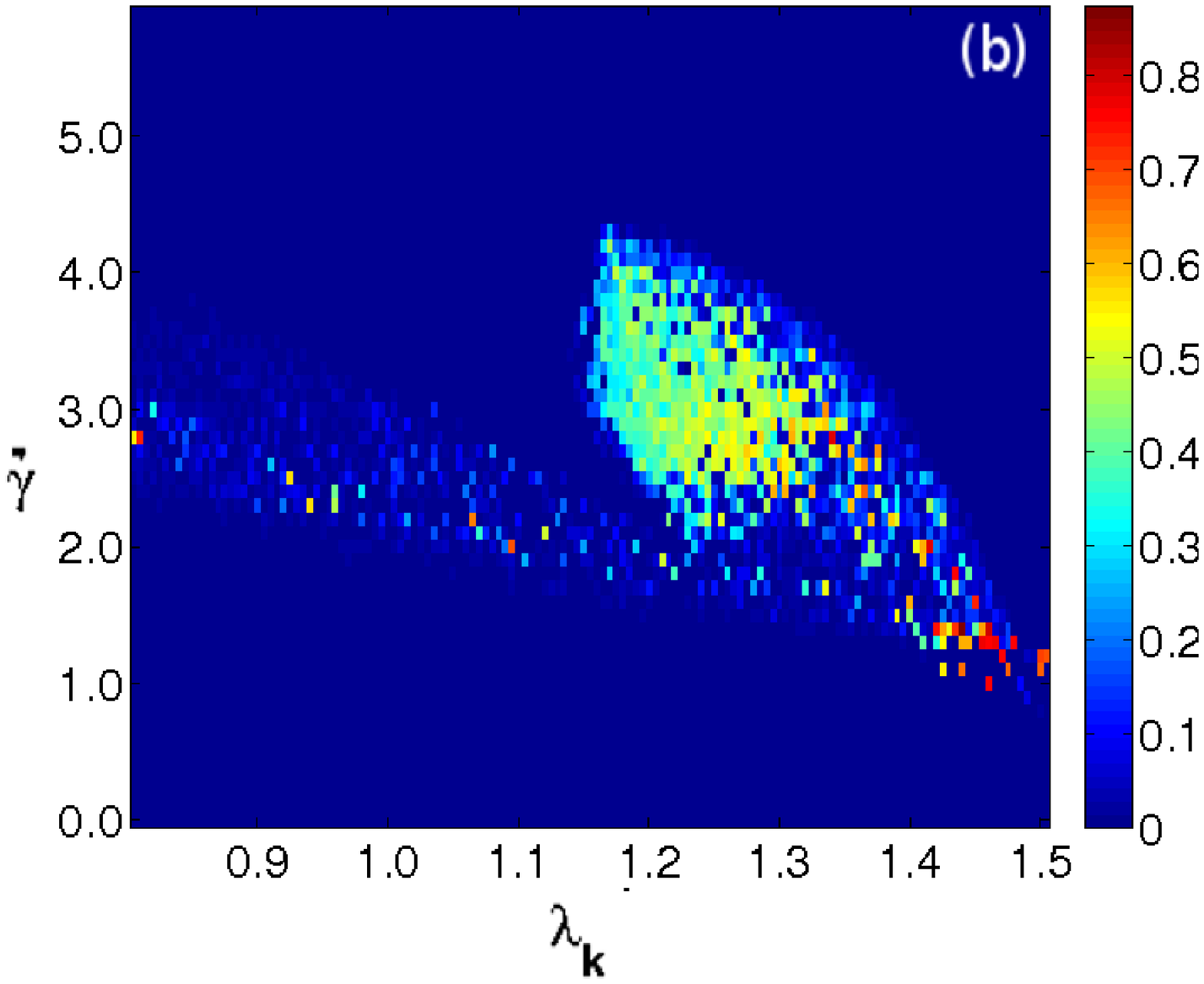}
\caption{[Color Online] Average deviation $\overline{d}$ (see text) from the mean value  of $a_0$ for
  $\epsilon$ = 0.1(right) and $\epsilon$ = 0.5(left). The quantity $\lambda_k$ is plotted on
  the $x$-axis and $\dot{\gamma}$on the $y$-axis. The lattice is a $100\times100$-site lattice.
  Note that this roughness is largest in the KT, KW and C regions, especially for large values of
  $\dot{\gamma}$. As $\epsilon$ is increased, the roughness decreases, as increasing spatial
  homogeneity is promoted.}
\label{av_dev}
\end{center}
\end{figure}

Fig.~(\ref{av_dev}) shows the space-time average of the deviation
defined in Eqn. \eqref{twod_dav}. The left panel (a) displays results for
$\epsilon = 0.1$ and the right panel (b) for $\epsilon = 0.5$. It is clear
that higher coupling strengths make the system more uniform in
space. Also, it appears that the regions with kayak-tumbling,
kayak-wagging and complex local dynamical behaviour show more
deviation in the spatial profile, exhibiting more spatial inhomogeneity.
\subsubsection{Regular regime}
In this section we discuss the spatial profile of our coupled map
lattice in two dimensions. As in the one-dimensional case, we start with random 
initial conditions and analyze the space profile after omitting a transient regime. 
We  analyse the density plots of the shear stress contribution to the order parameter,
in different dynamical regions. Considering 
the space-time behaviour of the system in the regular
region, with local dynamics belonging to the aligned, wagging,
tumbling and kayak-tumbling region, reveals spatially uniformity states which 
are periodicity in time. These are closely related to the states obtained in the oine-dimensional
case and are not discussed further here, as we will concentrate on results obtained 
in the physically more interesting C regime.
\subsubsection{Complex regime}
\begin{figure}
\begin{center}
\includegraphics[width=3.1in]{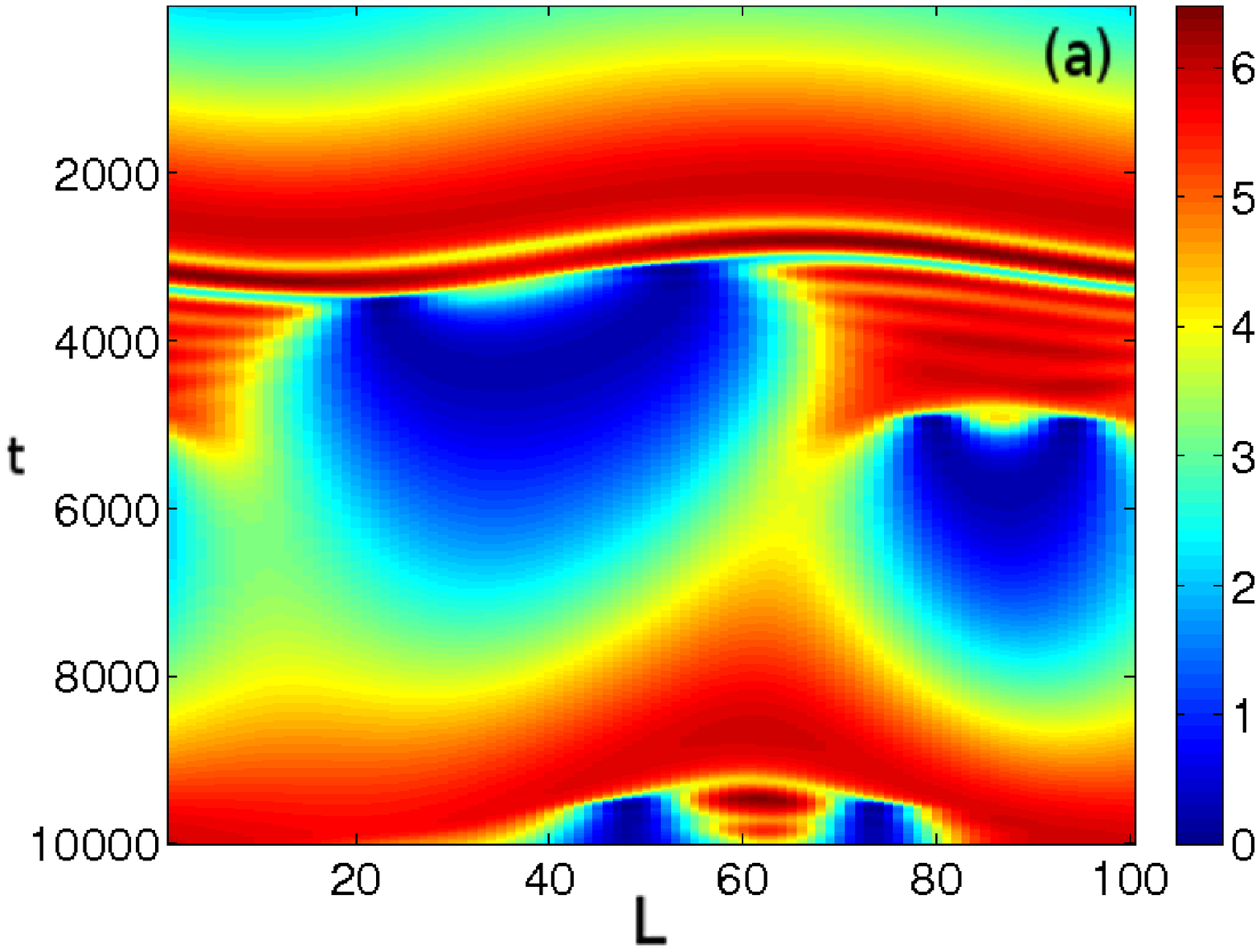}
\includegraphics[width=3.0in]{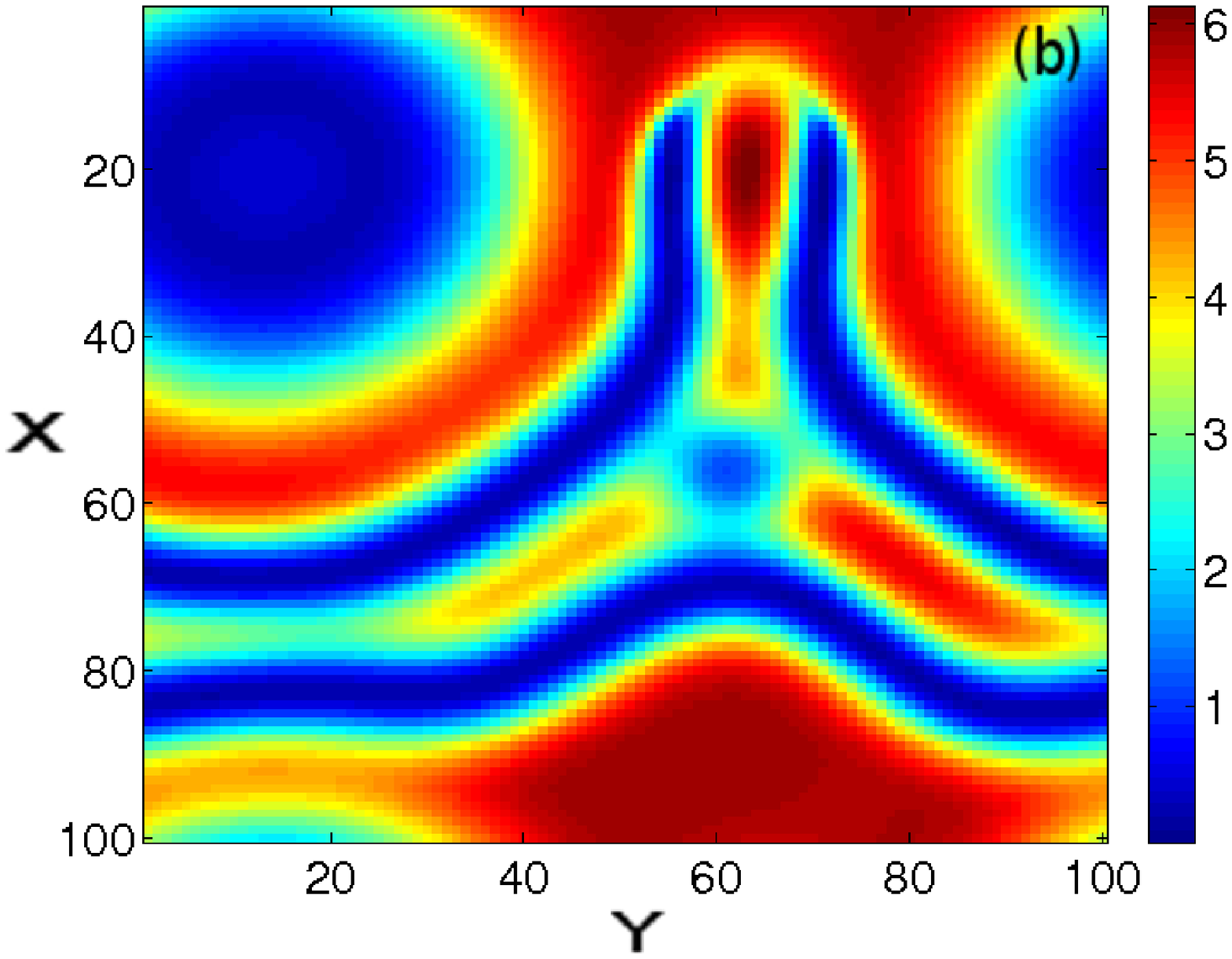}
\caption{[Color Online] (a) A color plot of order parameter part of shear stress. The quantities $\lambda_k$ = 1.25, $\dot{\gamma}$ =4.0 and $\epsilon = 0.1$.
  The time $t$ is plotted on the y-axis, which depicts the time evolution of the stress computed on 
  one row (x-axis) of the 100 $\times$ 100 lattice. (b) shows a snap shot of the full lattice at an intermediate
  time step. }
\label{stg4l1_25ep0_1}
\end{center}
\end{figure}

\begin{figure}
\begin{center}
\includegraphics[width=3.0in]{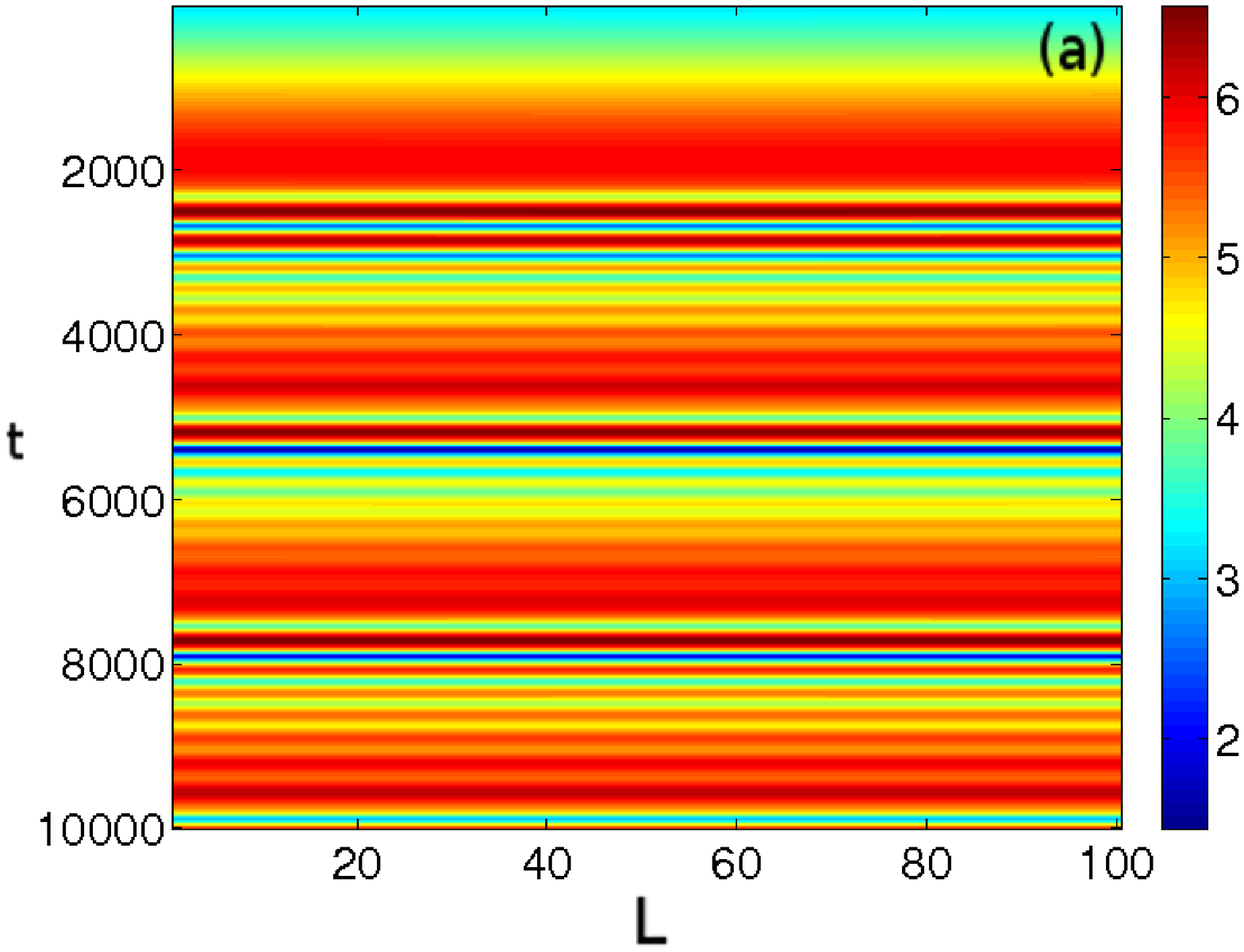}
\includegraphics[width=2.9in]{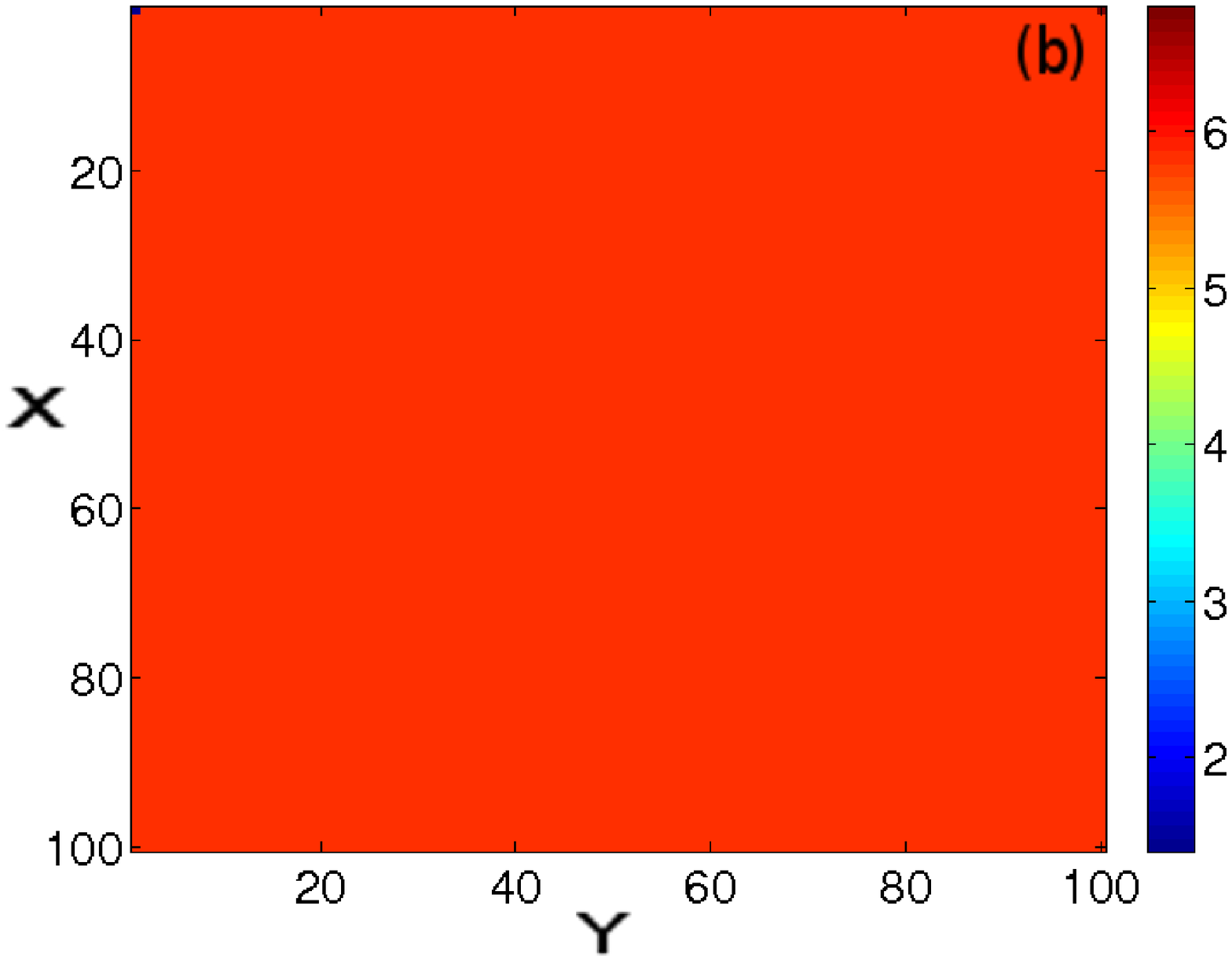}
\caption{[Color Online] (a) A color plot of order parameter part of shear stress. The quantities $\lambda_k$ = 1.25, $\dot{\gamma}$ =4.0 and $\epsilon = 0.5$.
  The time $t$ is plotted on the y-axis, which depicts the time evolution of the stress computed on 
  one row (x-axis) of the 100 $\times$ 100 lattice. (b) shows a snap shot of the full lattice at an intermediate
  time step. Note that increased spatial coupling favours the space uniform time-periodic state.}
\label{stg4l1_25ep0_5}
\end{center}
\end{figure}

\begin{figure}
\begin{center}
\includegraphics[width=5.0in]{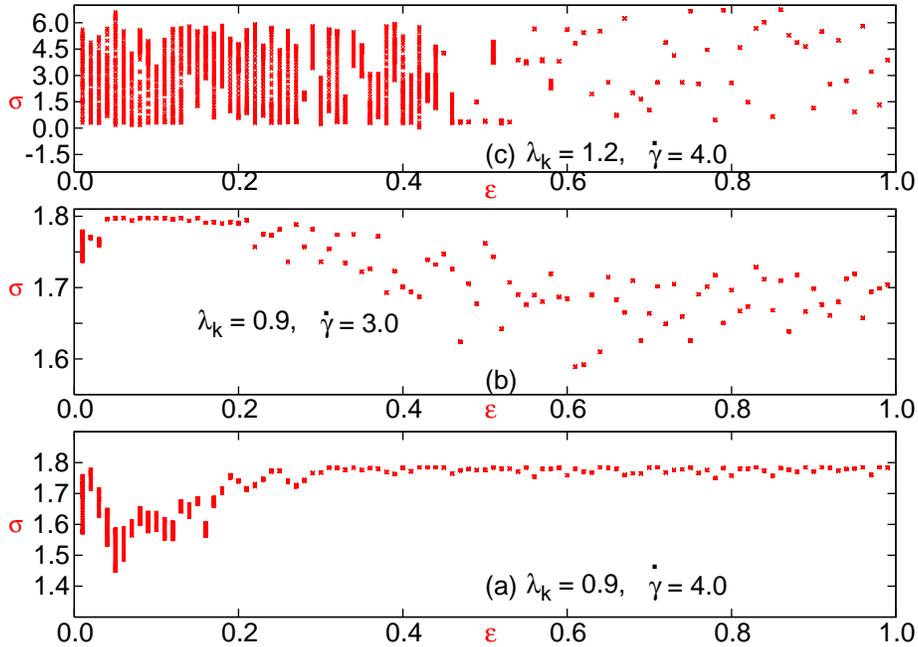}
\caption{[Color Online] Value of stress of one row of the lattice at one instant of
  time on the $y$ axis as a function of the  coupling strength $\epsilon$ on the $x$
  axis. Note that the broad spectrum of local stress values seen at small values of the
  coupling contract to an essentially unique value at large $\epsilon$.}
\label{coupling}
\end{center}
\end{figure}

The  configurations in Figs. (\ref{stg4l1_25ep0_1}) and (\ref{stg4l1_25ep0_5}) are from the complex region.
When the coupling becomes very large, one obtains spatially uniform states, as
clear from Figs.~ (\ref{stg4l1_25ep0_1}) and( \ref{stg4l1_25ep0_5}).
In Fig.~\ref{coupling}, we have chosen points (a) and (b) from the
KT region of the local phase diagram and (c) from the
C dynamical region. After leaving $10^4$ transient steps , we
have plotted one row of a $100 \times 100$ lattice at a single time
instant. On the $x$ axis we plot $\epsilon$ and on the $y$ axis we
plot the stress at $100$ points of the lattice at one time step. It is
evident that for high coupling strength $\epsilon$, the system goes to
a space-synchronized state. For low coupling constants, on the other
hand, there is a typically wide distribution of stress values at different
sites, indicating spatial inhomogeneity.


\section{Quantifying Spatio-Temporal Complexity}
\begin{figure}
\begin{center}
\includegraphics[width=6.0in]{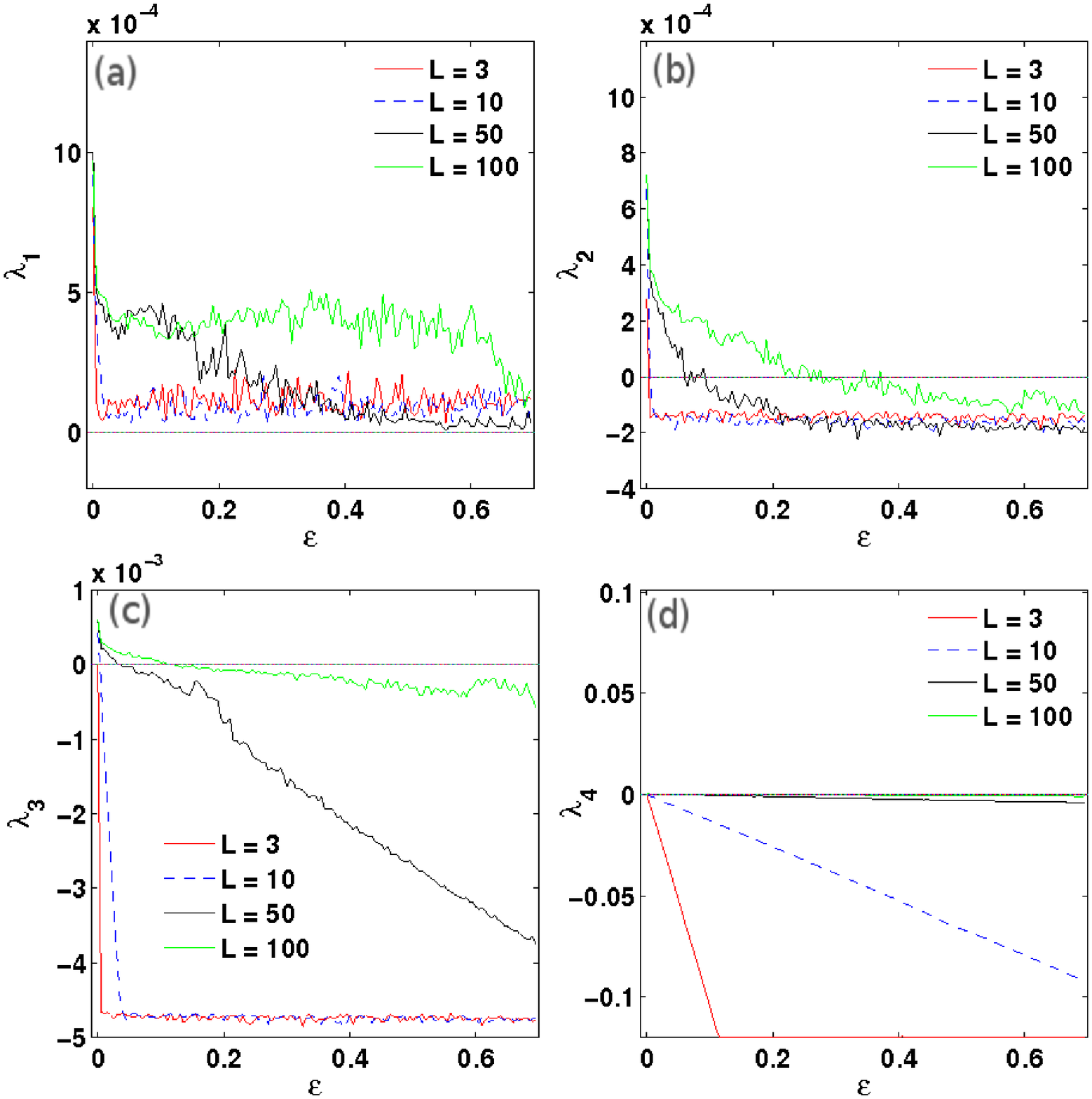}
\caption{[Color Online] Plots of the Lyapunov exponents obtained from our calculation (see text) for different
system sizes as a function of the coupling constant. The figures represent (a) he first Lyapunov
exponent $\lambda_1$ (b) the second Lyapunov exponent $\lambda_2$ c) the third Lyapunov
exponent $\lambda_3$ d) the fourth Lyapunov exponent $\lambda_4$. Note that all these exponents
tend to zero in the limit of large system sizes and that increasing the coupling between sites has the
effect of reducing the Lyapunov exponent if it were initially positive at zero coupling and increasing it,
if it were initially negative. }
\label{lyapunov}
\end{center}
\end{figure}

In this section, we report  results which quantify the spatio-temporal complexity in
the one-dimensional coupled map lattice specified in Eq.~\ref{local}. To understand
the nature of the complex behaviour represented in the phase diagram, we perform
calculations of the spectrum of Lyapunov exponents, as shown in Fig.~\ref{lyapunov}.
We first choose several values of the parameters
$\lambda_k$ and $\dot{\gamma}$ within the complex region and evolve the coupled
map. After waiting for an initial number of time steps to eliminate transients, we calculate
the Jacobian matrix  at each time step. We then consider a small deviation
from the attractor and iteratively multiply this deviation by the Jacobian, orthonormalizing
this vector at each time step. From this we  calculate the Lyapunov exponent,
using the method described in Ref.~\cite{lakshmanan}.

These results are illustrated in Fig.~\ref{lyapunov}, which exhibits the values of the
first four Lyapunov exponents, computed for parameter values 
$\lambda_k$ and $\dot{\gamma}$ for system sizes $L = 3, 10, 50$ and $100$,
as a function of the coupling constant $\epsilon$. As can be seen from these figures,
our results are the following. Qualitatively, in the complex regime, the first Lyapunov
exponent is always positive, even as the system size and the spatial coupling are increased.
The local value of this exponent is also positive. This
value decreases further with spatial coupling but remains positive. Roughly speaking,
larger lattice sizes show larger values for this exponent, consistent with results from the one-dimensional
PDE calculation. These values appear to
saturate for small coupling values but decrease for larger values of the spatial coupling.

The second and higher order Lyapunov exponents, in our calculation, are small and {\em negative} 
for the smallest lattice sizes, but move to  values that are close to zero as the lattice size is 
increased. At small couplings, for the larger lattices,  this value is positive but goes negative as the coupling strength
is increased. Thus, the data for the Lyapunov exponents are consistent with the general
conclusion that going to larger lattice sizes stabilizes chaos, whereas increasing 
the coupling between sites suppresses complex spatial behaviour. 
The clustering of Lyapunov exponents around zero in the large
system size limit is consistent with the emergence of spatio-temporal intermittency on large scales \cite{chate1993,anantha,neelima}.

%==============================================================================================%
\section{Conclusions}

In summary, this paper reports a study of a coupled map lattice model
constructed to study spatio-temporal aspects of rheological chaos in sheared
nematic solutions. Our study was based on the construction of a suitable local
map capable of reproducing  the physics of the spatially uncoupled (equivalently, uniform) 
limit, including the variety of phases and the complex phase diagram obtained for that case. 
When such maps are placed on a regular (one or two-dimensional) lattice and coupled diffusively
through a variety of coupling schemes, the local dynamics in the coupled map shares close similarities 
with the map describing the spatially  uncoupled case. 

Our studies of the coupled map
in both one and two dimensions indicates that regimes of regular behaviour largely exhibit space-uniform
and time-periodic states, with the coupled dynamics roughly following the uncoupled case. We have
analysed the dynamical behaviour of the two quantities which characterize local order in the nematic,
the uniaxial $s_1$ and the biaxial $s_2$ order parameters, examining their time evolution in the
different states.

In contrast, in  the complex or $C$ region of the local phase diagram, such coupling leads to states that exhibit 
spatio-temporal intermittency and chaos. We have characterized such states by examining  the Lyapunov spectra
as well as the frequency 
dependence of the time series of physical quantities such as the stress. We find evidence for a broad, 
power-law distribution of time-scales in the problem. Further, in the complex region, one  often sees a coexistence of 
regular (lamina) and chaotic regimes as a prelude to fully developed chaos in which dynamical fluctuations occur
independently from site to site. In some regimes,  periodic bands immersed in a more complex,
fluctuating background are obtained,  suggestive of the possibility of transient shear bands stabilized by the 
dynamics, a feature also present in ODE-based studies of this problem\cite{sr,sr1,debarshini}. 
The basic scale of these complex dynamical  patterns is
alterable by changing the coupling constant, indicative of self-similarity in the spatio-temporally intermittent case. 
At very large values of the coupling constant, the
space profile is expected to become uniform; however, for small and intermediate values of this
coupling constant, the spectrum of Lyapunov exponents merges to zero,  consistent with our
observation of generic spatio-temporal intermittency in the weak coupling case.

We have experimented with using spatial coupling terms which represent the advective effects of the
shear flow, coupled to fixed boundary conditions where the orientation and magnitude of the order
parameter are fixed at the boundary. Such terms appear, at small amplitude, to mainly distort the 
sorts of dynamical structures obtained for the symmetric coupling state and seem to evolve smoothly
from them.

The usefulness of coupled map lattice representations of the spatio-temporal dynamics of
systems exhibiting chaos in their local dynamics is that such representations often provide
both useful physical insights as well as are computationally easier to simulate than their
PDE versions.   In that sense, the problem of rheochaos in sheared nematics offers an ideal
setting for CML methods, since the {\em local} dynamics of the sheared nematic is highly non-trivial,
exhibiting a variety of temporally periodic as well as chaotic states. 
As shown here,  the variety of non-trivial spatio-temporal behaviour exhibited by sheared nematics is very largely 
a consequence of simply coupling these dynamical degrees of freedom in space. The physics appears
substantially independent of how precisely this spatial coupling is done, with  the simple lattice model with parallel
update exhibiting virtually all the behaviour of the more complex and computationally intensive
studies of the appropriate PDE's. This, together with the specific results presented in this paper for our
coupled map approach to rheochaos in sheared nematics, is our central conclusion.

Further, order-parameter-based models, such as the one described in this paper and in the work of 
Refs.~\cite{debarshini, sr,sr1}, contain essential non-linear terms in the free energy. It is these terms that are
responsible for the non-trivial local dynamics captured in our local map as well as in the coupled
map lattice. Ref.~\cite{debarshini} emphasizes the role of  ``additional complex collective dynamics'' 
arising from such  nonlinearities which is not captured in the DLS model but is relevant to the
qualitative nature of the intermittent and chaotic behaviour seen in this system. Such non-linearities
are naturally accounted for in our approach.

Our study of the spatio-temporal dynamics of sheared nematics using CML
methods possibly represents the first extension of such methods to the problem of rheochaos. In
contrast to previous work based on ODE's which studied only the one-dimensional case, it is 
relatively easy to extend our CML methodology to higher dimensions, even to the experimentally  
relevant three-dimensional case.  It would be  interesting to see how, if at all, hydrodynamic 
effects can be incorporated in models  of this type. Whether other experimental
systems of sheared complex fluids which exhibiting rheochaos can be fruitfully analysed  
using similar coupled map approaches remains to be seen.

\acknowledgements GIM acknowledges partial support from DST, India. The authors acknowledge
the PRISM project at IMSc for  assistance and useful conversations with A.K. Sood, C. Dasgupta,
S. Ramaswamy and M.E. Cates.


\begin{thebibliography}{}
\bibitem{footnote} We have explored alternative constructions
for such a local map in Ref.~\cite{kamil}, where we studied a quaternion representation
of the local orientational degrees of freedom. Incorporating spatial coupling is,
however, easiest in the version studied in this paper.
\end{thebibliography}
\end{document}